# Signal-to-noise and spatial resolution in in-line imaging. 3. Optimization using a simple model

**Timur E. Gureyev[a]\*, David M. Paganin[b] and Harry M. Quiney[a]**

[a] School of Physics, University of Melbourne, Parkville, Victoria, 3010, Australia

[b] School of Physics and Astronomy, Monash University, Clayton, Victoria, 3800, Australia

Correspondence email: timur.gureyev@unimelb.edu.au

**Funding information**     National Health and Medical Research Council (grant No. APP2011204).

**Synopsis**  Geometrical magnification and X-ray energy are optimized with respect to spatial resolution, contrast, noise, radiation dose and other characteristics of propagation-based phase-contrast X-ray imaging setups.

**Abstract**    The problem of optimization of propagation-based phase-contrast imaging setups is considered in the case of projection X-ray imaging and three-dimensional tomography with phase retrieval. For two-dimensional imaging, a simple model for a homogeneous edge feature embedded in a bulk sample is used to obtain analytical expressions for the image intensity. This model allows for explicit optimization of the geometrical parameters of the imaging setup and the choice of X-ray energy that maximizes the image contrast or the contrast-to-noise ratio. We also consider the question of optimization of the biomedical X-ray imaging quality characteristic which balances the contrast-to-noise against the spatial resolution and the radiation dose. In the three-dimensional case corresponding to propagation-based phase-contrast tomography with phase retrieval according to Paganin's method, the optimization of the imaging setup is studied with respect to the source size, the detector resolution, the geometrical magnification and the X-ray energy.

## 1. Introduction

Propagation-based phase-contrast imaging (PBI) and tomography (PB-CT) have been shown to deliver superior image contrast and contrast-to-noise (CNR) compared to conventional attenuation-based imaging and CT at the same radiation dose and spatial resolution when imaging low-Z materials using hard X-rays (Paganin, 2006; Wilkins *et al.*, 2014; Endrizzi, 2018; Quenot *et al.*, 2022). After about 30 years of active development towards beneficial applications in medical and biomedical imaging, this technology is finally approaching the stage where it can soon be used to image live humans at radiation doses comparable with, or lower than, conventional X-ray absorption-based methods. In order to make practical implementations of the PBI and PB-CT imaging technologies as effective as possible, it is essential to find the optimal parameters for the corresponding imaging setups. Note that a microfocus source is typically required in PBI imaging to provide an X-ray beam with sufficient spatial coherence (Wilkins *et al.*, 1996). The main issue with such sources at present is the trade-off between the need to reduce the effective size of the region emitting X-rays in order to deliver the required spatial coherence and the need for the source to be sufficiently bright to enable the acquisition of a planar image or a CT scan within a reasonable time. This time can typically be of the order of 10-15 seconds during which the patient could be reasonably expected to be able to hold their breath. The spatial resolution of the detector needs to be considered alongside the X-ray source size for determining the spatial resolution in the images.

The geometry of a PBI imaging setup includes a number of key parameters that must be included into any optimization process. One such parameter is the source-to-sample distance that affects the X-ray flux, the maximum illuminated area and the penumbral blurring due to the finite source size. Another key parameter is the sample-to-detector distance, which needs to be sufficiently large in order to allow the propagation-based phase contrast (Snigirev *et al.*, 1995) to become sufficiently strong to guarantee adequate signal-to-noise ratio (SNR) and CNR in the images. At the same time, the sample-to-detector distance, together with the source-to-sample distance, determines the geometric magnification of the imaging setup which affects image quality via the interplay with the spatial resolution of the detector (Gureyev *et al.,* 2008). The quality of PBI images usually improves linearly with increasing source-to-detector distance, but that distance is typically the subject of practical constraints imposed by the size of the premises where the X-ray scanner can be hosted. It is also important to consider the optimization of the X-ray energy or, more generally, the X-ray spectrum produced by the X-ray sources and possibly modified by suitable filters and monochromators, that would maximize the PBI image quality at a given radiation dose.

In view of the above considerations, it is clearly important to decide at the start what characteristics of the PBI image should be optimized for practical purposes, such as design of laboratory-based microfocus X-ray scanners or synchrotron-based setups. Obvious candidates for such characteristics are the SNR/CNR, the spatial resolution, the Detective Quantum Efficiency (DQE) and the radiation dose (Bezak et al., 2021). More recently, we introduced and studied additional image quality characteristics, such as the intrinsic imaging quality and the biomedical X-ray imaging quality (Gureyev *et al.*, 2014, 2020, 2025). The latter characteristics combine the SNR/CNR, spatial resolution and the radiation dose into single metrics that are invariant with respect to linear image filtering (such as e.g. detector pixel binning) and provide quantitative measures of the information channel capacity of the imaging system per single incident photon (Gureyev *et al.*, 2016). However, while these image quality metrics can certainly be helpful in the context of biomedical X-ray imaging applications, the ultimate benchmark for a medical imaging instrument is its diagnostic performance (Barrett & Myers, 2004). This involves assessments of collected images by medical imaging specialists, such as radiologists (Longo *et al.*, 2017; Taba *et al.*, 2020). The problem of correlation between the "objective" image quality characteristics, such as CNR and spatial resolution, and the "subjective" evaluation of the quality of the same images by medical imaging specialists, has been researched in the context of PBI (Baran *et al.*, 2017; Tavakoli Taba *et al.*, 2019). While some correlations between the subjective and objective image quality characteristics in PBI have been reliably established, this question still remains at least partially contentious overall. In the present study, we only address the objective image quality characteristics. A comparison with the optimization of the subjective image quality of PBI setups can be the subject of a future study.

Regarding the previously published literature on closely related topics, apart from the references given above, we would like to mention, in particular, the papers (Nesterets *et al.*, 2005; Gureyev *et al.*, 2008; Brombal *et al.*, 2018; Nesterets *et al.*, 2018; Delogu *et al.*, 2019; Oliva *et al.*, 2020). The work described in (Nesterets *et al.*, 2005) was based on a generalized weak-object approximation and reported optimization results for contrast, CNR and spatial resolution in PBI. In (Gureyev *et al.*, 2008), the results of analytical study of the PBI contrast, SNR and spatial resolution were reported as functions of the same geometric parameters of the imaging setup as discussed above. This study was based on a simple "toy" model of a pure phase (non-absorbing) edge feature imaged in PBI settings. It was established that the SNR and contrast produced by such a pure-phase edge feature initially increased linearly with the effective propagation distance in the near-Fresnel region and then asymptoted to a constant value at longer distances. The characteristic behavior of the spatial resolution was opposite in the sense that it remained approximately constant in the near-Fresnel region and then, at further distances (i.e. for smaller Fresnel numbers (Hecht, 2017)), it increased linearly with the effective propagation distance, in proportion to the width of the first Fresnel zone (the width of the first Fresnel fringe in the image of the edge (Hecht, 2017)). In (Brombal *et al.*, 2018), the effect of the propagation distance on spatial resolution, contrast and SNR was investigated both theoretically and experimentally. Experimental and numerical optimization of the X-ray energy in synchrotron-based imaging of breast tissue was studied in detail in (Delogu *et al.*, 2019; Oliva *et al.*, 2020). The publication (Nesterets *et al.*, 2018) contained results that are largely complementary to those reported below. While in the present work we partially follow in the footsteps of (Gureyev *et* *al.*, 2008) by using a simple model for the imaged sample, the PBI configurations studied in (Nesterets *et al.*, 2018) were more general and detailed, perhaps, at the expense of simplicity. The latter results included, for example, optimization conditions for PBI setups using X-ray spectra similar to those produced by real solid-anode sources and realistic detector point-spread functions (PSFs). In contrast, in the present paper we optimize the X-ray energy explicitly only in the monochromatic case which is more relevant to synchrotron imaging. Correspondingly, we use the imaging parameters typical to those of a synchrotron beamline, such as the Imaging and Medical Beamline (IMBL) of the Australian Synchrotron (Stevenson *et al.*, 2017), in our numerical examples. However, we show that the optimization of the geometric parameters of PBI setups can usually be performed independently of the X-ray wavelength, which opens the way for performing the geometric optimization at multiple wavelengths separately and then simply integrating the results over the relevant X-ray spectrum. In the present work, we also use a simple "homogenenous" weakly-absorbing edge model, which generalizes the non-absorbing edge model utilized in (Gureyev *et al.*, 2008). This approach allows us to apply Paganin's homogeneous Transport of Intensity (TIE-Hom) method of phase retrieval in PBI and PB-CT (Paganin *et al.*, 2002; Paganin, 2006). We also study for the first time the problem of PBI optimization with respect to the biomedical X-ray imaging quality characteristic (Gureyev *et al.*, 2025), which should make our results particularly useful for the design of future medical PBI and PB-CT imaging instruments.

**2. PBI contrast produced by an embedded monomorphous edge**

Let a sample be located immediately before the "object" plane $z = 0$ transverse to the optical axis $z$, and $(x, y)$ be the Cartesian coordinates in the transverse planes (Fig. 1). The sample is illuminated by an X-ray beam emanating from a small spatially incoherent source located near the point $z = -R_1$. The sample consists of a uniform "bulk" material and an embedded "edge feature" (Fig. 1). Let $n_0(z, \lambda) = 1 - \delta_0(z, \lambda) + i\beta_0(z, \lambda)$ be the complex refractive index of the "bulk" material, where $\lambda$ is the X-ray wavelength, $n_0(z, \lambda) = 0$ outside the "bulk" slab, $-T_0 \leq z < -T$, and is uniform within that slab. The complex refractive index of the edge feature, $n_1(x, z, \lambda) = 1 - \delta_1(x, \lambda) + i\beta_1(x, \lambda)$, is equal to zero outside a smaller slab, $-T \leq z \leq 0$, $T << T_0$, is uniform in the *y* direction within that slab, and has a shape of a smooth edge increasing in density along the *x* direction (Fig. 1). Furthermore, the difference between the two refractive indexes inside the edge slab, $-T \leq z \leq 0$, is assumed to be monomorphous, in the sense that $\delta(x, z, \lambda) = \gamma(\lambda)\beta(x, z, \lambda)$ for all points inside the slab, where $\delta(x, z, \lambda) = \delta_1(x, z, \lambda) - \delta_0(z, \lambda)$, $\beta(x, z, \lambda) = \beta_1(x, z, \lambda) - \beta_0(z, \lambda)$, and the proportionality coefficient $\gamma(\lambda)$ is constant within the slab (Paganin *et al.*, 2002; Paganin, 2006). The complex refractive index outside the whole sample slab, $-T_0 \leq z \leq 0$, is equal to unity (corresponding to vacuum). We also assume that the sample is thin, in the sense that $T_0 \ll \min(R_1, R_2)$, in which case the exact z-location of the thin edge within the sample does not matter.

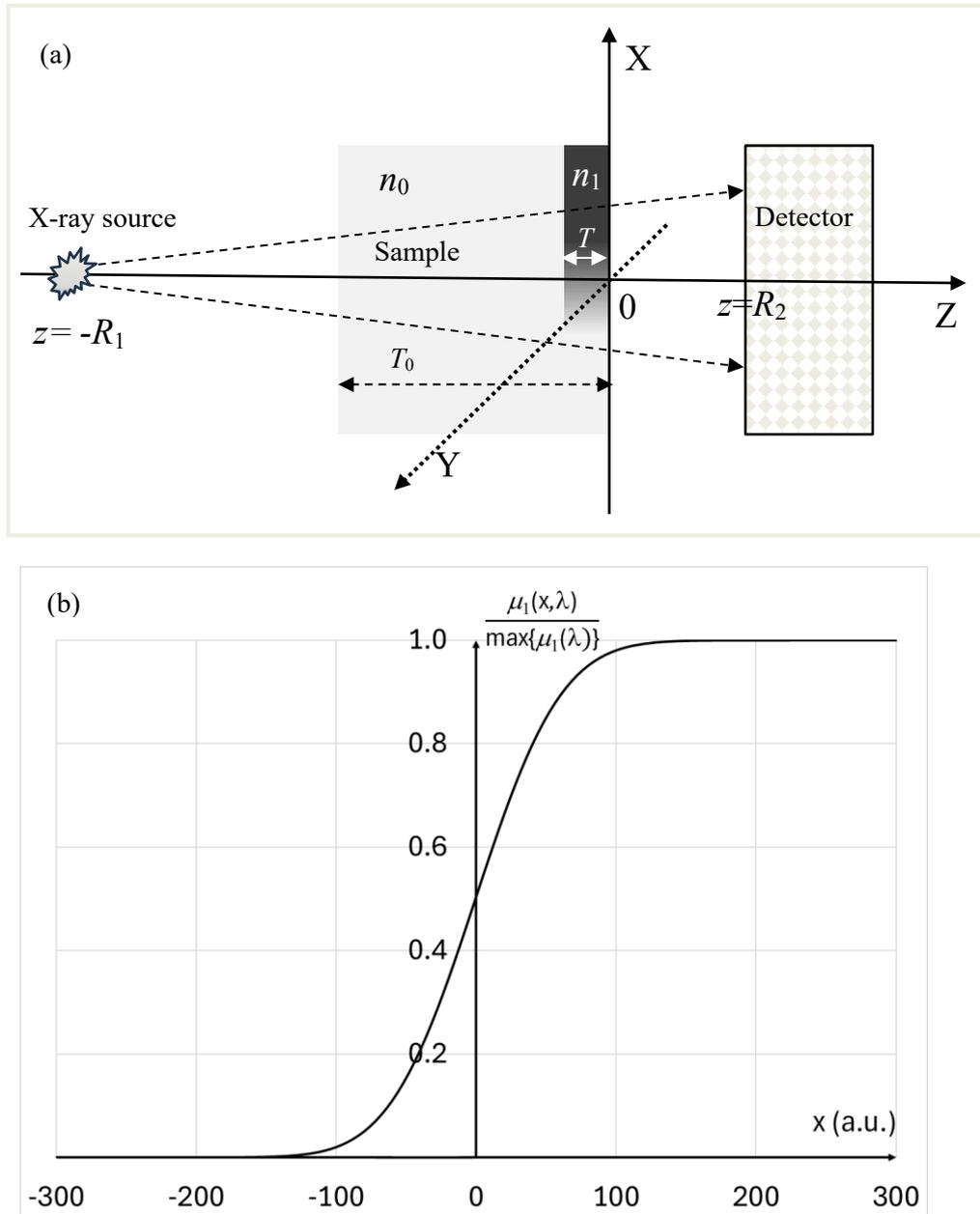

**Figure 1** (a) Setup of propagation-based X-ray imaging of a uniform "bulk" sample containing a monomorphous "edge" feature. Both the bulk sample and the edge are assumed to be uniformly extended along the Y axis. (b) X-profile of the linear attenuation coefficient of the edge feature.

The X-ray transmission through the sample can be characterized by the complex transmission function $\exp(ikT_0)\exp[i\varphi_0(\lambda) - B_0(\lambda)/2]\exp[i\varphi(x,\lambda) - B(x,\lambda)/2]$, where
$\varphi_0(\lambda) = (2\pi/\lambda)\delta_0(\lambda)T_0$, $B_0(\lambda) = (4\pi/\lambda)\beta_0(\lambda)T_0 = \mu_0(\lambda)T_0$, $\varphi(x,\lambda) = (2\pi/\lambda)\int_0^T \delta(x,z,\lambda)dz$
and $B(x,\lambda) = (4\pi/\lambda)\int_0^T \beta(x,z,\lambda)dz = \int_0^T \mu(x,z,\lambda)dz$. The assumption of the monomorphicity made above implies that $\varphi(x,\lambda) = \gamma(\lambda)B(x,\lambda)/2$.

The transmitted beam is registered by a position-sensitive detector located immediately after the "detector" plane $z = R_2$. The X-ray transmission profile of the edge feature, $\exp[-B(x,\lambda)]$, is defined by the maximum absorption $(\mu T)_{max}(\lambda) = \mu(+\infty, 0, \lambda)T > 0$ and the "shape function" $E(x; \sigma_{obj})$:

$$B(x,\lambda) = (\mu T)_{max}(\lambda) E(x; \sigma_{obj}), \quad E(x; \sigma_{obj}) = H(x) * G(x; \sigma_{obj}), \tag{1}$$

where the asterisk denotes one-dimensional convolution, $H(x)$ is the Heaviside "step" function (which is equal to 0 for negative and zero $x$, and equal to 1 for positive $x$) and $G(x, \sigma_{obj})$ is a Gaussian function, $G(x,\sigma) = (2\pi\sigma^2)^{-1/2}\exp[-x^2/(2\sigma^2)]$, with the standard deviation $\sigma_{obj}$ describing the "intrinsic unsharpness" ("blurriness") of the edge. Note that in this case the function $E(x, \sigma_{obj})$ is a cumulative Gaussian distribution:

$$E(x,\sigma) = \int_{-\infty}^{x} G(x';\sigma)dx' = (1/2)\{1 + erf[x/(\sqrt{2}\sigma)]\}, \tag{2}$$

where $erf(x) = (2/\sqrt{\pi})\int_0^x \exp(-t^2)dt$ is the error function. Note also that $E(-\infty, \sigma) = 0$ and $E(+\infty, \sigma) = 1$. Similar edge models were used previously, for example, in (Nesterets *et* al., 2005; Gureyev *et al.*, 2008; Aloo *et al.*, 2022).

The transmitted X-ray photon fluence (expressed as photons per unit area) (Barrett & Myers, 2004) in the vicinity of the edge feature in the object plane $z = 0$ can be modeled as:

$$I(x, y, 0, \lambda) = I_{id}(x, \lambda) * G(x, \sigma_{sys}(1)), \tag{3}$$

where $I_{id}(x,\lambda) \equiv I_{in}(\lambda)\exp[-B_0(\lambda) - B(x,\lambda)]$ is the transmitted photon fluence in the object plane in the case of an ideal imaging system with delta-function LSF, $I_{in}(\lambda)$ is the photon fluence of the incident beam and $G(x, \sigma_{sys}(M))$ is the line-spread function (LSF) of the in-line imaging system. Note that we have assumed that the source-to-object distance $R_1$ is much larger than the characteristic dimensions of the edge feature and, therefore, it is possible to neglect the dependence of the incident photon fluence $I_{in}(\lambda)$ on the transverse spatial coordinates $(x, y)$. The LSF is assumed to be Gaussian (we also assume for simplicity that the LSF is the same at all X-ray energies), with variance $\sigma_{sys}^2(M) = (M-1)^2 M^{-2} \sigma_{src}^2 + M^{-2} \sigma_{det}^2$, where $\sigma_{src}$ and $\sigma_{det}$ are the standard deviations of the source intensity distribution and the detector LSF, respectively, and $M = (R_1 + R_2)/R_1$ is the geometric magnification (Gureyev et al., 2008). This form of $\sigma_{sys}(M)$ is a direct consequence of the projection imaging geometry (Fig. 1). At the two extreme values of $M$, we have $\sigma_{sys}(1) = \sigma_{det}$ and $\sigma_{sys}(\infty) = \sigma_{src}$. It is straightforward to verify (Nesterets *et al.*, 2005) that the minimal possible $\sigma_{sys}(M)$ is achieved at $M = M_{res} \equiv 1 + (\sigma_{det}^2 / \sigma_{src}^2)$ and is equal to

$$\sigma_{sys}(M_{res}) = \frac{\sigma_{src} \sigma_{det}}{\sqrt{\sigma_{src}^2 + \sigma_{det}^2}} = \frac{\sigma_{src}\sqrt{M_{res}-1}}{\sqrt{M_{res}}} = \frac{\sigma_{det}}{\sqrt{M_{res}}}. \qquad (4)$$

When $\sigma_{src} = \sigma_{det}$, we have $M_{res} = 2$ and $\sigma_{sys}(2) = \sigma_{src}/\sqrt{2} = \sigma_{det}/\sqrt{2}$. At magnification $M = M_{res}$, the spatial resolution in in-line imaging is always finer than both the source and the detector resolutions. In order to properly assess the spatial resolution in the acquired images, however, it is not enough to just consider the geometric magnification, but it is also necessary to take into account the effect of free-space propagation (Fresnel diffraction).

It is well known that, at sufficiently short propagation distances $z$, the spatial distribution of the photon fluence, $I(x, y, z, \lambda)$, in in-line images can be described by the Transport of Intensity equation (TIE) (Teague, 1983; Paganin, 2006). As we are considering a one-dimensional edge-like feature that is uniform along the $y$ coordinate, all image intensity distributions will be constant along $y$, and therefore we will omit the coordinate $y$ from the notation below for brevity. Substituting eq.(3) into the monochromatic TIE-Hom (Paganin *et al.*, 2002), we obtain in the image plane $z = R_2$:

$$I(Mx, R_2, \lambda) = M^{-2}(1 - a^2 \partial_{xx}^2) I_{id}(x, \lambda) * G(x, \sigma_{sys}(M)), \qquad (5)$$

where $I(x, R_2, \lambda)$ is the photon fluence distribution in the image plane $z = R_2$, $a^2 = \gamma R' \lambda / (4\pi)$ and $R' = R_2 / M$ is the effective propagation ("defocus") distance. Equation (5) can be expanded as

$$I(Mx, R_2, \lambda) = M^{-2} I_{in}(\lambda) \exp[-B_0(\lambda)] \times$$
$$\{\exp[-B(x,\lambda)] * G(x, \sigma_{sys}) - a^2 \partial_{xx}^2 \exp[-B(x,\lambda)] * G(x, \sigma_{sys})\}. \quad (6)$$

Note that $\partial_x H(x) = \delta_D(x)$, where $\delta_D(x)$ is the Dirac delta function. Therefore,

$$\partial_x \exp[-B(x,\lambda)] = -\exp[-B(x,\lambda)](\mu T)_{max}(\lambda) \delta_D(x) * G(x, \sigma_{obj})$$

$= -\exp[-B(x,\lambda)](\mu T)_{max}(\lambda) G(x, \sigma_{obj})$. The additivity of variance in the convolution of Gaussian functions implies that $G(x, \sigma_{obj}) * G(x, \sigma_{sys}) = G(x, \sigma_M)$, where

$\sigma_M^2 = \sigma_{sys}^2(M) + \sigma_{obj}^2 = (M-1)^2 M^{-2} \sigma_{src}^2 + M^{-2} \sigma_{det}^2 + \sigma_{obj}^2$. Using this and the fact that, according to the validity conditions of eq.(5), $B(x, \lambda)$ must be slowly varying (Gureyev et al., 2008), we obtain

$\partial_x \{\partial_x \exp[-B(x,\lambda)] * G(x, \sigma_{sys})\} \cong \exp[-B(x,\lambda)](\mu T)_{max} G(x, \sigma_M) x / \sigma_M^2$. Taking this relationship into account, we can re-write eq.(6) as

$$I(Mx, R_2, \lambda) = M^{-2} I_{in}(\lambda) \exp[-B_0(\lambda)] \times$$
$$\{\exp[-B(x,\lambda)] * G(x, \sigma_{sys}) - (\gamma / N_F)(\mu T)_{max}(\lambda) \exp[-B(x,\lambda)] x G(x, \sigma_M)\}, \quad (7)$$

where $\gamma / N_F = \gamma R' \lambda / (4\pi \sigma_M^2) = a^2 / \sigma_M^2$, and $N_F = \Delta_M^2 / R'\lambda$ is the "minimal Fresnel number" corresponding to the characteristic width, $\Delta_M \equiv 2\sqrt{\pi}\sigma_M$, of the image of the edge (see Fig. 2). Equation (7) describes the evolution of the photon fluence in the vicinity of the image of the monomorphous edge as a function of propagation distance and other parameters of the imaging setup. The term $\exp[-B(x,\lambda)] * G(x, \sigma_{sys})$ in eq.(7) corresponds to absorption contrast. It depends on the propagation distance only via the change in the blurring of the edge with the magnification $M$. The second term inside the curly brackets in eq.(7), $(\gamma / N_F)(\mu T)_{max}(\lambda) \exp[-B(x,\lambda)] x G(x, \sigma_M)$, corresponds to phase contrast. The phase-contrast term also changes its width as a function of magnification. However, unlike the absorption term, the phase term's amplitude increases with the effective propagation distance $R'$ (see Fig. 2).

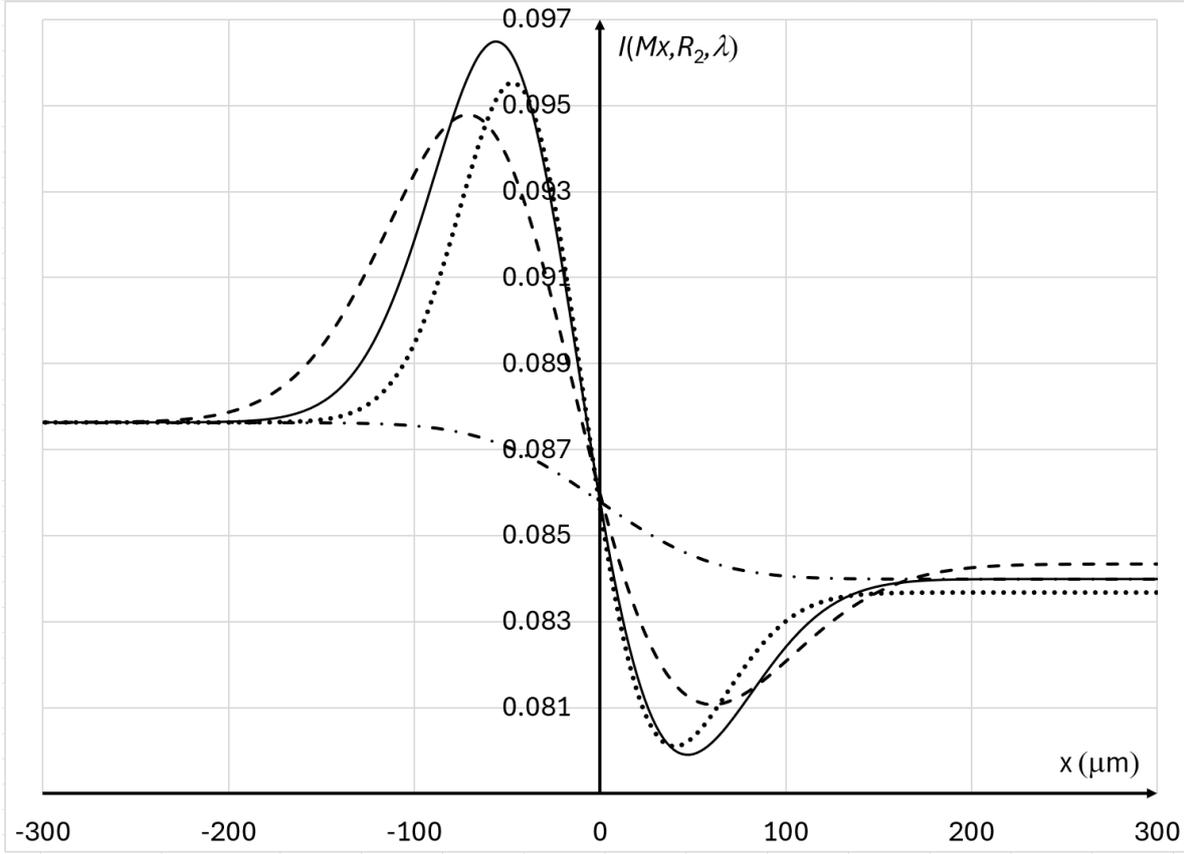

**Figure 2** PBI intensity profiles, expressed by eq.(7), in three cases corresponding to existing and proposed configurations for imaging breast tissue specimens at the IMBL beamline of the Australian Synchrotron (see Table 1), with $\lambda = 0.3875$ Å ($E = 32$ keV), $I_{in}(\lambda) = 1$, and different magnifications: $M = 1.05$ ($R_2 \cong 6.67$ m, dotted line), $M = 1.094$ ($R_2 \cong 12.0$ m, solid line), $M = 1.15$ ($R_2 \cong 18.3$ m, dashed line). The dot-dashed line shows the absorption component only from eq.(7) in the case $M = 1.094$. The profiles for $M = 1.05$ and $M = 1.15$ have been shifted vertically to bring the left ends (which correspond to the absence of the edge feature) to the same fluence level as in the case $M = 1.094$, in order to facilitate visual comparison of the contrasts.

Note that the TIE-Hom equation in general and eq.(7) in particular are valid only in the so-called near-Fresnel regime, which imposes an upper limit on the magnitude of the phase contrast. Indeed, a sufficient condition for the near-Fresnel regime in the present setup is $N_{F,obj} \gg (\mu T)_{max} \gamma$, where $N_{F,obj} \equiv \Delta_{obj}^2 / R'\lambda$ and $\Delta_{obj} \equiv 2\sqrt{\pi}\sigma_{obj}$ (Gureyev et al., 2008). Since, $N_F \geq N_{F,obj}$, it implies that $N_F \gg (\mu T)_{max} \gamma$, or

$$\gamma / N_F \ll 1/(\mu T)_{max}. \tag{8}$$

It is easy to verify that $|xG(x,\sigma_M)| \leq 1/\sqrt{2\pi e}$, and therefore eq.(8) implies that the phase term in eq.(7) is always much smaller than $(1/\sqrt{8\pi e})$. Note, however, that when $(\mu T)_{max} \ll 1$ (in the weak absorption case), it is still possible to have $\gamma/N_F \gg 1$, implying that the phase-contrast term in eq.(7) can be much larger than the absorption-contrast term $1-\exp[-B(x,\lambda)] \cong B(x,\lambda) \leq (\mu T)_{max} \ll 1$.

Let us apply the approach that was previously employed in (Gureyev et al., 2008) for calculating the image contrast for an edge-like feature in PBI. In that approach, the "propagation contrast" was associated with the difference in image intensity at points $x = \mp \sigma_M$, i.e. approximately at the maximum and minimum of the first Fresnel fringe (see Fig. 2). Using eq.(7) and approximating $\exp[-B(-\sigma_M,\lambda)] \cong 1$, $\exp[-B(\sigma_M,\lambda)] \cong \exp[-(\mu T)_{max}(\lambda)]$, the propagation contrast can be expressed as

$$C = \frac{I(-M\sigma_M, R_2, \lambda) - I(M\sigma_M, R_2, \lambda)}{I(-M\sigma_M, R_2, \lambda) + I(M\sigma_M, R_2, \lambda)} =$$

$$\frac{[1-q_{min}(\lambda)] + [1+q_{min}(\lambda)](2\pi e)^{-1/2}(\gamma/N_F)(\mu T)_{max}(\lambda)q_{min}(\lambda)}{[1+q_{min}(\lambda)] + [1-q_{min}(\lambda)](2\pi e)^{-1/2}(\gamma/N_F)(\mu T)_{max}(\lambda)q_{min}(\lambda)},$$

where $q_{min}(\lambda) = \exp[-(\mu T)_{max}(\lambda)]$ is the minimal transmission of the edge feature. Using the constraint from eq.(8), we can neglect the second additive term in the denominator of the last expression and obtain:

$$C(M,\lambda) \cong \frac{1-q_{min}(\lambda)}{1+q_{min}(\lambda)} + \frac{(\mu T)_{max}(\lambda) q_{min}(\lambda)}{(2\pi e)^{1/2}} \frac{\gamma}{N_F}. \tag{9}$$

The first additive term (fraction) in eq.(9) corresponds to absorption contrast, while the second additive term corresponds to phase contrast. It follows from eq.(8) that the phase-contrast term must be small in the near-Fresnel region. However, when the absorption contrast is small, $(\mu T)_{max} \ll 1$, the phase-contrast term can be much larger than the absorption-contrast term, since it is possible to have $\gamma/N_F \gg 1$, as noted earlier.

## 3. Optimization of in-line phase contrast and CNR

In the numerical simulations used for verification of theoretical results in this paper, we will use the parameters shown in Tables 1 and 2, which roughly correspond to current and prospective setups for imaging breast tissue samples at IMBL.

**Table 1** Geometrical parameters of the imaging setup used in the numerical simulations. The notation for all included quantities is explained in the main text of the paper. Magnification values different from the ones shown in this table are also used in the text.

| R(m) | $\sigma_{det}$(μm) | $\sigma_{src}$(μm) | $T_0$(cm) | T(cm) | M | $\sigma_{sys}$(μm) | R'(m) |
|---|---|---|---|---|---|---|---|
| 140.0 | 37.5 | 400 | 8.60 | 0.50 | 1.05 | 40.5 | 6.35 |
| | | | | | 1.094 | 48.5 | 11.0 |
| | | | | | 1.15 | 61.5 | 15.9 |

**Table 2** X-ray energy (wavelength) related parameters of the imaging setup used in the numerical simulations. The notation for all included quantities is explained in the main text of the paper.

| E(keV) | λ(Å) | $\mu_0$(μm$^{-1}$) | μ(μm$^{-1}$) | γ |
|---|---|---|---|---|
| 32.0 | 0.3875 | 2.62E-05 | 8.50E-06 | 869 |
| 26.0 | 0.4769 | 3.35E-05 | 1.42E-05 | 642 |
| 42.0 | 0.2952 | 2.15E-05 | 4.67E-06 | 1203 |

Let us consider imaging conditions that maximize the phase contrast in eq.(9). Apart from the constant factor $2^{-5/2}\pi^{-3/2}e^{-1/2}$, the phase-contrast CNR can be represented as a product of two distinct terms, $R'/\sigma_M^2$ and $\gamma(\lambda)\lambda(\mu T)_{max}(\lambda)\exp[-(\mu T)_{max}(\lambda)]$, the first one being a function of the geometrical parameters of the imaging setup and the second one depending on the X-ray wavelength. Therefore, it is logical to consider two separate problems: (A) maximization of the term $R'/\sigma_M^2$ with respect to the source-to-sample and sample-to-detector distances, and (B) maximization of the term $\gamma(\lambda)\lambda(\mu T)_{max}(\lambda)\exp[-(\mu T)_{max}(\lambda)]$ with respect to the X-ray energy.

Regarding problem (A), we will consider the case where the total source-to-detector distance $R = R_1 + R_2$ is fixed and the edge is sharp, in the sense that the "intrinsic unsharpness" $\sigma_{obj}$ of the edge can be neglected, i.e. $\sigma_M^2 \cong \sigma_{sys}^2(M)$. The expression $R'/\sigma_{sys}^2(M)$ needs to be maximized as a function of magnification $M = (R_1 + R_2)/R_1$. Expressing $R' = R(M-1)/M^2$ and $R'/\sigma_{sys}^2 = R(M-1)/[(M-1)^2\sigma_{src}^2 + \sigma_{det}^2]$, it is easy to check that the equation $d(R'/\sigma_{sys}^2)/d(M-1) = 0$ has the solution $M_C = 1 + \sigma_{det}/\sigma_{src}$. It has a well-known special case of $M_C = M_{res} = 2$, when $\sigma_{det} = \sigma_{src}$. Note that at the optimal magnification, the source and the detector always make equal contributions to the contrast, because $\sigma_{src}(M_C - 1) = \sigma_{det}$. Also, $R'/\sigma_{sys}^2(M_C) = R/(2\sigma_{det}\sigma_{src})$. Therefore, the propagation contrast produced by a sharp monomorphous edge at the optimal magnification $M_C$ is equal to

$$C(M_C, \lambda) \cong \frac{1 - q_{min}(\lambda)}{1 + q_{min}(\lambda)} + \frac{\gamma R \lambda (\mu T)_{max}(\lambda) q_{min}(\lambda)}{(8\pi e)^{1/2} \Delta_{src} \Delta_{det}}, \quad M_C = 1 + \frac{\sigma_{det}}{\sigma_{src}}, \tag{10}$$

where $\Delta_{src} = 2\sqrt{\pi}\sigma_{src}$ and $\Delta_{det} = 2\sqrt{\pi}\sigma_{det}$ are the widths of the source and detector components of the PSF, respectively (Gureyev *et al.*, 2024, 2025). The phase contrast in eq.(10) is linearly proportional to the total source-to-detector distance and is inversely proportional to both the source size and the detector resolution. Note that eq.(10) does not include the case of a parallel-beam geometry. However, it can be easily verified that the PBI in a parallel-beam geometry can be formally obtained by setting $\Delta_{src} = \Delta_{det}/2$ in eq.(10). In the case of setups corresponding to Tables 1 and 2, the magnification maximizing the contrast is equal to $M_C = 1 + 75\,\mu m/800\,\mu m \cong 1.094$, which corresponds to sample-to-detector distance $R_2 \cong 12$ m.

For problem (B), we need to consider the dependence of the expression $f(\lambda) = \gamma(\lambda)\lambda(\mu T)_{max}(\lambda)\exp[-(\mu T)_{max}(\lambda)]$ on $\lambda$. Away from X-ray absorption edges, we have (see e.g. Gureyev *et al.*, 2001): $\mu(\lambda) = (4\pi/\lambda)\beta(\lambda) \cong \mu(\lambda_0)(\lambda/\lambda_0)^3$, $\beta(\lambda) \cong \beta(\lambda_0)(\lambda/\lambda_0)^4$, $\delta(\lambda) \cong \delta(\lambda_0)(\lambda/\lambda_0)^2$, where $\lambda_0$ is an arbitrary value within a chosen suitably limited interval of wavelengths. Let us introduce a temporary notation $\gamma(\lambda)\lambda \cong \gamma(\lambda_0)\lambda_0(\lambda/\lambda_0)^{-1} = a\lambda^{-1}$ and $(\mu T)_{max}(\lambda) = (\mu T)_{max}(\lambda_0)(\lambda/\lambda_0)^3 = b\lambda^3$. In this notation, $f(\lambda) = ab\lambda^2\exp(-b\lambda^3)$. The equation $df(\lambda)/d\lambda = ab\lambda(2 - 3b\lambda^3)\exp(-b\lambda^3) = 0$ has a root $\lambda_C = [2/(3b)]^{1/3}$, which corresponds to the maximum $f(\lambda_C) = (2/3)e^{-2/3}\gamma(\lambda_C)\lambda_C$. In practice, the optimal wavelength $\lambda_C$ can be found experimentally from the condition $q_{min}(\lambda_C) \equiv \exp[-(\mu T)_{max}(\lambda_C)] = e^{-2/3} \cong 0.51$, corresponding to the requirement that the mean X-ray transmission through the edge feature should be around 51%. The optimal contrast at this wavelength is equal to

$$C(M, \lambda_C) \cong a_0 + c_0 \frac{\gamma(\lambda_C)\lambda_C R'}{\Delta_M^2}, \quad \exp[-(\mu T)_{max}(\lambda_C)] = e^{-2/3}, \tag{11}$$

where $a_0 \equiv (1-e^{-2/3})/(1+e^{-2/3}) \cong 0.322$, $c_0 \equiv (2/3)e^{-2/3}(2\pi e)^{-1/2} \cong 0.083$ and $\Delta_M = 2\sqrt{\pi}\sigma_M$.

Figure 3 shows the profiles of the detected X-ray fluence near the edge feature, calculated in accordance with eq.(7) at three different X-ray energies in the setup described by Tables 1 and 2. Note that in the case of parameters from Tables 1 and 2 the optimal energy maximising the contrast of the edge feature is approximately 12 keV ($\lambda \cong 1.03$ Å). However, the X-ray transmission through the bulk of the sample at such low energy will be extremely low, and therefore the noise level will be very high (see the discussion below).

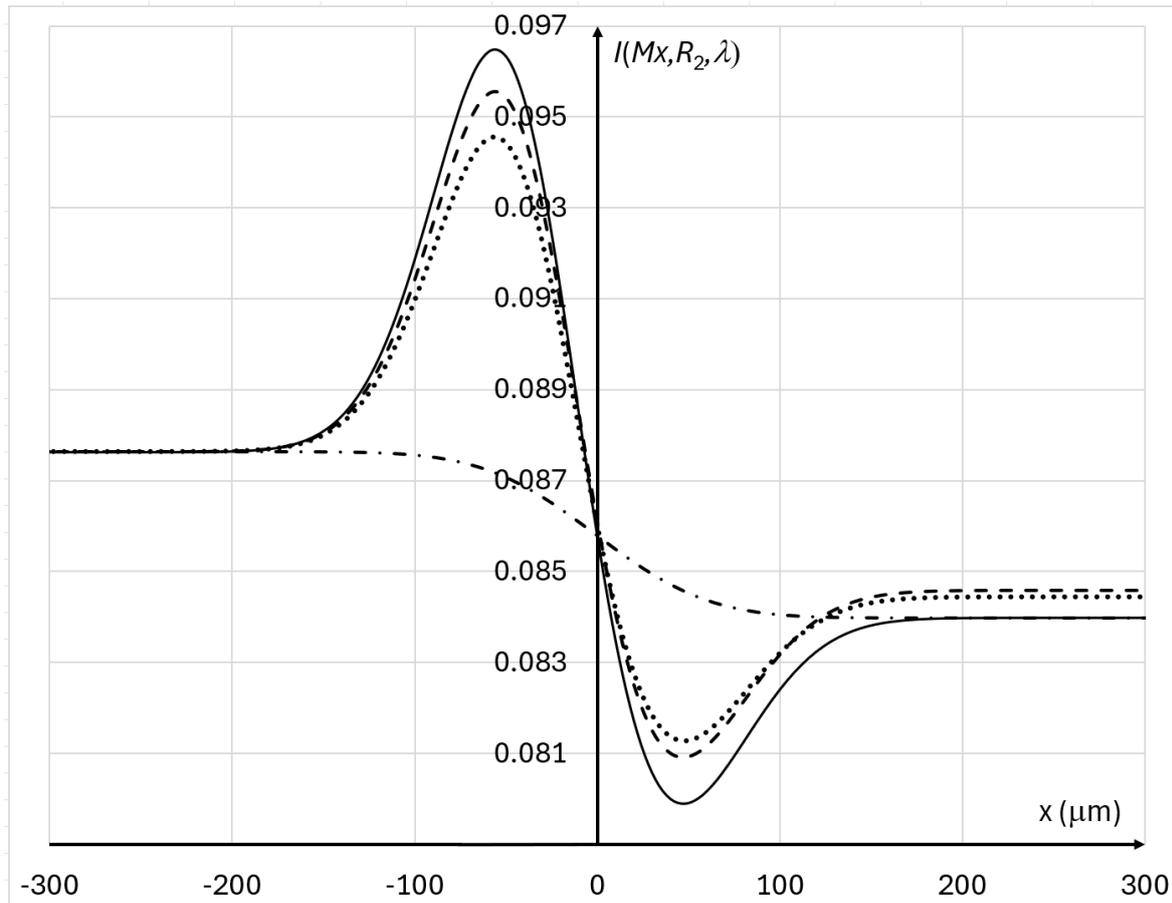

**Figure 3** PBI intensity profiles, expressed by eq.(7), in the cases corresponding to some proposed configurations for imaging breast tissue specimens at the IMBL beamline of the Australian Synchrotron (see Tables 1 and 2), with $M = 1.094$ ($R_2 \cong 12.0$ m), $I_{in}(\lambda) = 1$, and different X-ray wavelengths (energies): $\lambda = 0.4769$ Å ($E = 26$ keV, dotted line), $\lambda = 0.3875$ Å ($E = 32$ keV, solid line), $\lambda = 0.2952$ Å ($E = 42$ keV, dashed line). The dot-dashed line shows the absorption component only from eq.(7) in the case of $E = 32$ keV. The profiles for $E = 26$ keV and $E = 42$ keV have been shifted vertically to bring the left ends (which correspond to the absence of the edge feature) to the same fluence level as in the case $E = 32$ keV, in order to facilitate visual comparison of the contrasts.

Finally, if both the magnification and the X-ray wavelength are optimized in the case of a sharp edge, the maximum contrast becomes:

$$C(M_C, \lambda_C) \cong a_0 + \frac{c_0}{2} \frac{\gamma(\lambda_C)\lambda_C R}{\Delta_{src} \Delta_{det}}. \qquad (12)$$

As noted above, this maximum possible value of the propagation contrast is achieved at the X-ray wavelength $\lambda_C$, at which (i) the minimal X-ray transmission through the edge feature is around 51% ($\exp[-(\mu T)_{max}(\lambda_C)] = e^{-2/3}$), and (ii) the magnification is equal to $M_C = 1 + \sigma_{det}/\sigma_{src}$. As mentioned earlier in conjunction with eq.(10), the second (phase contrast) term in in eqs.(11) and (12) cannot be larger than unity, because of the validity conditions imposed by eq.(8).

Note however that the optimization of the contrast with respect to the X-ray energy considered above is not very realistic: it favours strong absorption in the feature, without properly taking into account the effect of absorption in the bulk of the object. This happens because the term in eq.(7) that corresponds to the bulk absorption, $\exp[-B_0(\lambda)]$, cancels out in the expression for the contrast, eq.(9). Therefore, although the contrast produced by the edge feature at high X-ray absorption may formally be strong, the fact that only a few photons get through the bulk of the sample is going to adversely affect the quality of the corresponding image. A related image quality characteristic that adequately accounts for this phenomenon is the contrast-to-noise ratio (CNR).

We define CNR as the product of the contrast and the SNR. In order to evaluate the SNR in propagation images of a monomorphous edge, we assume that the photon counting statistics is Poissonian (Barrett & Myers, 2004). Then the average squared SNR of the photon fluence can be expressed via the incident fluence, $I_{in}(\lambda)$, as $\text{SNR}^2(\lambda) = \eta M^{-2} I_{in}(\lambda) \exp[-B_0(\lambda)]\Delta_{det}^2$, where $\eta$ is the quantum efficiency of the detector. Combining this with eq.(9), we obtain the following expression for the CNR:

$$\text{CNR}(M,\lambda) \cong \eta^{1/2} M^{-1} I_{in}^{1/2}(\lambda) \exp[-\mu_0(\lambda)T_0/2]\Delta_{det} \times$$
$$\left[ \frac{1-q_{min}(\lambda)}{1+q_{min}(\lambda)} + (2\pi e)^{-1/2}(\gamma/N_F)(\mu T)_{max}(\lambda)q_{min}(\lambda) \right]. \qquad (13)$$

Proceeding exactly as in the case of the optimization of contrast with respect to $M$, we obtain that the magnification maximizing the phase-contrast part of the CNR, $M_{CNR}$, in the case of a sharp edge, satisfies the equation $d[R'/(M\sigma_{sys}^2)]/d(M-1) = 0$. This leads to a cubic equation $2(M_{CNR}-1)^3 + (M_{CNR}-1)^2 - \sigma_{det}^2/\sigma_{src}^2 = 0$ for $M_{CNR}$. Although the roots of this equation can be expressed analytically in terms of the ratio $\sigma_{det}^2/\sigma_{src}^2$ using Cardano's formula, the corresponding expressions are cumbersome and thus not very useful. In practice, one can find the roots of this equation for any given numerical value of $\sigma_{det}/\sigma_{src}$ using, for example, Wolfram Mathematica (Wolfram Research Inc., 2025). We obtained by this method that $M_{CNR} \cong 1.657$ in the case $\sigma_{src} = \sigma_{det}$, while in the case corresponding to the IMBL imaging setup parameters in Tables 1 and 2, the positive root of the cubic equation is $M_{CNR} \cong 1.087$ ($R_2 \cong 11.2$ m).

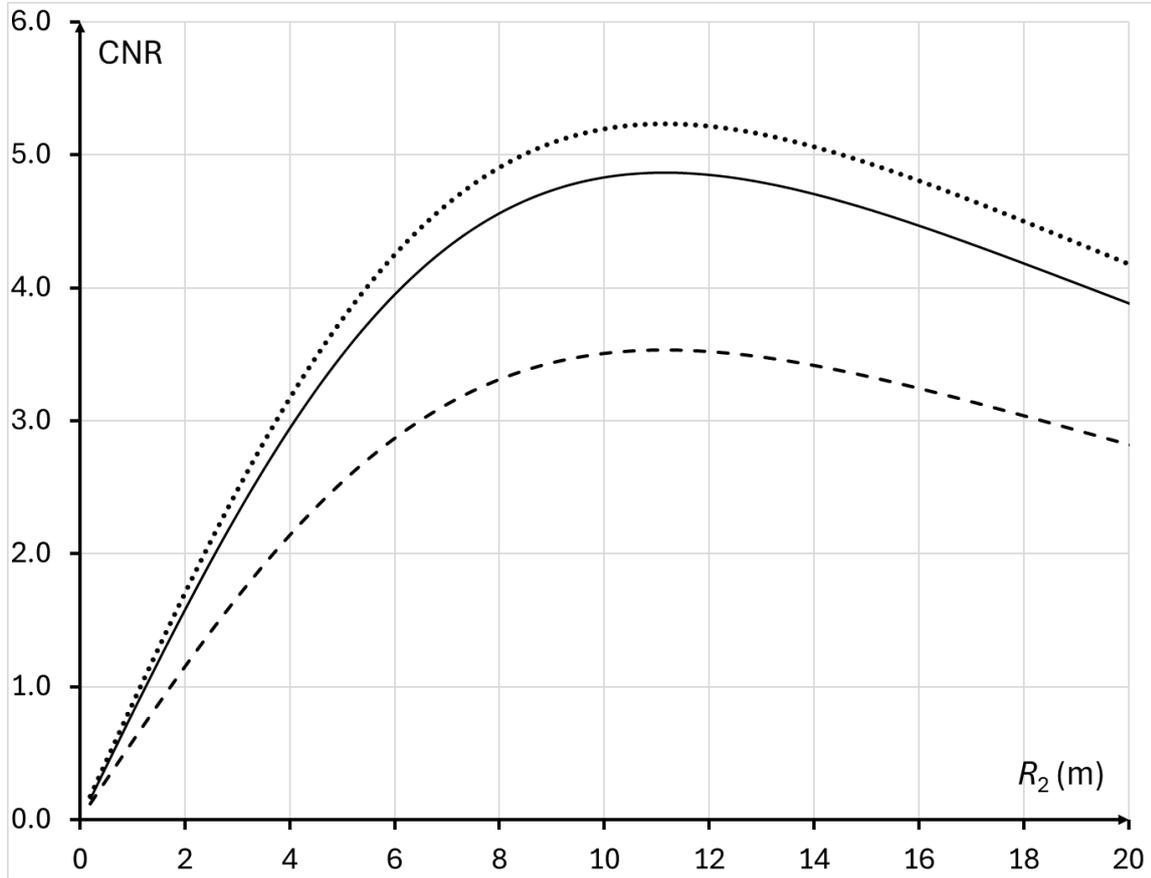

**Figure 4** CNR, as expressed by eq.(13), in the setups with parameters from Tables 1 and 2, 0.1 m ≤ $R_2$ ≤ 20 m, $I_{in}(\lambda) = 1$ μm$^{-2}$, and different X-ray wavelengths (energies): $\lambda = 0.4769$ Å ($E = 26$ keV, dotted line), $\lambda = 0.3875$ Å ($E = 32$ keV, solid line), $\lambda = 0.2952$ Å ($E = 42$ keV, dashed line). The optimal magnification in this case is equal to $M_{CNR} \cong 1.087$ ($R_2 \cong 11.2$ m).

This result agrees with direct numerical evaluation of eq.(15) presented in Fig. 4 for the imaging setup corresponding to Tables 1 and 2. Alternatively, it is easy to rewrite the above cubic equation in the form $\sigma_{det}/\sigma_{src} = (M_{CNR}-1)(2M_{CNR}-1)^{1/2}$, which allows one to create a look-up table or a graph, with a one-to-one correspondence between the optimal magnification values and the corresponding ratios of the detector resolution to the source size (see the solid line in Fig. 5). It is easy to see from Fig. 5 that $M_{CNR} < M_C = 1 + \Delta_{det}/\Delta_{src}$ for all values of $\Delta_{det}/\Delta_{src}$.

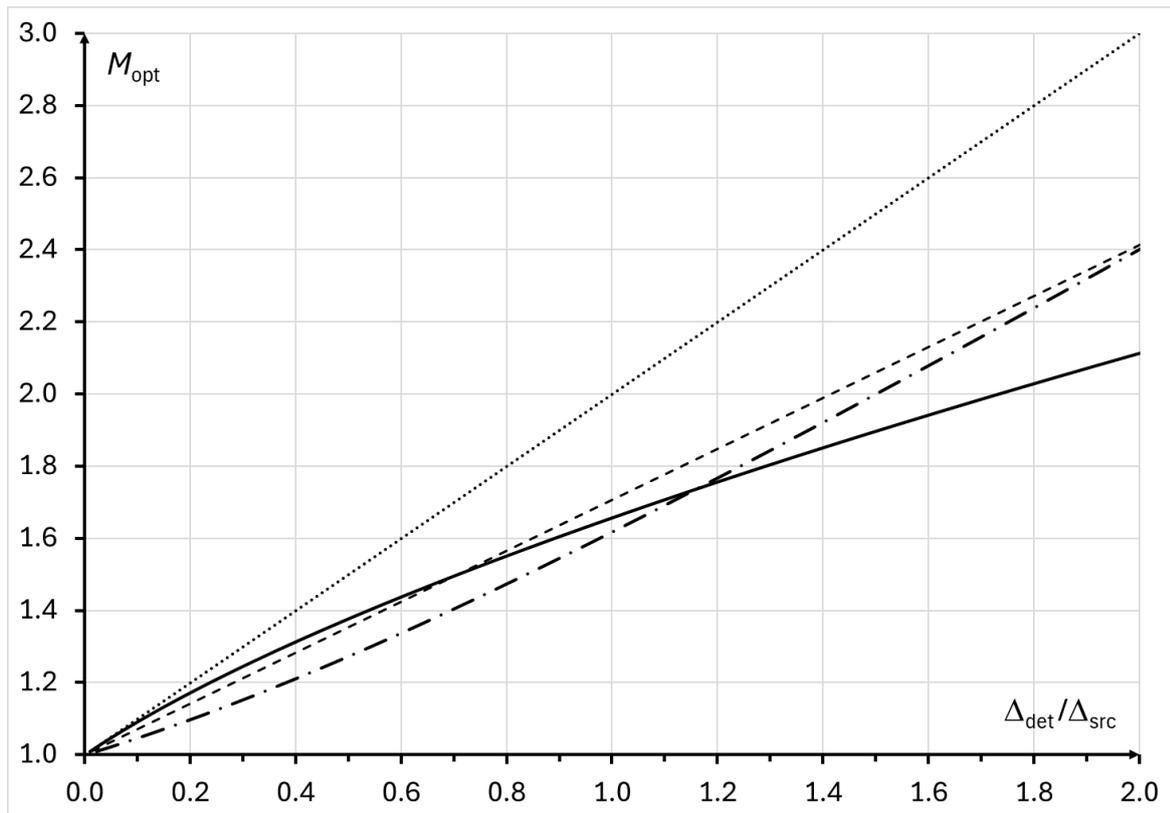

**Figure 5** Optimal magnification $M_{opt}$ as a function of the ratio of the detector resolution to the X-ray source size, $\Delta_{det}/\Delta_{src}$ ($=\sigma_{det}/\sigma_{src}$), in the cases of: 1) PBI contrast, $M_{opt} = M_C = 1 + \Delta_{det}/\Delta_{src}$ (dotted line); 2) CNR, $M_{opt} = M_{CNR}$ (solid line); 3) biomedical X-ray imaging quality in 2D PBI images, $M_{opt} = M_{Q2} = 1 + \Delta_{det}/(\sqrt{2}\,\Delta_{src})$ (dashed line); 4) biomedical X-ray imaging quality in PB-CT reconstructions: $M_{opt} = M_{Q3}$ (dash-dotted line) ($M_{Q2}$ and $M_{Q3}$ are defined in Section 4 below).

Optimization of the CNR with respect to the X-ray wavelength leads to the same equations as in the case of image contrast considered above, with the only difference that instead of the term $\exp[-(\mu T)_{max}(\lambda)]$ corresponding to X-ray absorption in the edge feature in the case of contrast, in the case of CNR we get $\exp[-\mu_0(\lambda)T_0/2-(\mu T)_{\max}(\lambda)]$. However, the tabulated values of $\mu_0$ for the materials of interest (adipose breast tissue) within the energy (wavelength) interval of most interest for breast imaging, i.e. approximately $20\,\text{keV} \leq E \leq 40\,\text{keV}$, indicate that, unlike the cubic behaviour in the case of $\mu(\lambda)=(4\pi/\lambda)\beta(\lambda)\cong\mu(\lambda_0)(\lambda/\lambda_0)^3$, $\mu_0(\lambda)$ is almost linear with respect to the wavelength: $\mu_0(\lambda)\cong\mu_0(\lambda_0)(\lambda/\lambda_0)$, see Fig. 6 (NIST, 2025; TS-Imaging, 2025). In the case of breast tissue, this fact was also investigated in a recent experimental study (Soares *et al.*, 2020). We hypothesize that the $\lambda$-linear terms in the expressions for $\mu_0(\lambda)$ and $\mu_1(\lambda)$ largely cancel each other in the expression for $\mu(\lambda)=\mu_1(\lambda)-\mu_0(\lambda)$, leaving the cubic terms as the dominant ones. Recall also that the term $\exp[-\mu_0(\lambda)T_0]$ corresponds to the X-ray absorption in the bulk of the sample. When the edge feature is small compared to the bulk object (which is the case frequently encountered in practice), the X-ray transmission at the optimum wavelength, $\lambda_{CNR}$, is determined primarily by the absorption in the bulk of the sample, rather than in the edge feature. Let us use the previously introduced notation $\gamma(\lambda)\lambda \cong \gamma(\lambda_0)\lambda_0(\lambda/\lambda_0)^{-1} = a\lambda^{-1}$ and $(\mu T)_{max}(\lambda) = (\mu T)_{max}(\lambda_0)(\lambda/\lambda_0)^3 = b\lambda^3$, and combine it with the modified $\lambda$-dependence of $\mu_0(\lambda)$: $\mu_0(\lambda) \cong \mu_0(\lambda_0)(\lambda/\lambda_0) = c\lambda$. In this notation, the phase-contrast part of the CNR in eq.(13) in the case of a small edge feature can be expressed as $f(\lambda) = ab\lambda^2 \exp(-c\lambda)$. The equation $df(\lambda)/d\lambda = ab\lambda(2 - c\lambda)\exp(-c\lambda) = 0$ has a root $\lambda_{CNR} = 2/c$, which corresponds to the maximum $f(\lambda_{\mathrm{CNR}}) = e^{-2}\gamma(\lambda_{\mathrm{CNR}})\lambda_{\mathrm{CNR}}(\mu T)_{\max}(\lambda_{\mathrm{CNR}})$. In practice, the optimal wavelength $\lambda_{\mathrm{CNR}}$ can be found experimentally from the condition $\exp[-\mu_0(\lambda_{\mathrm{CNR}})T_0] = e^{-4} \cong 0.02$, corresponding to the requirement that the mean X-ray transmission through the sample should be around 2%. Examples of the dependencies of the phase-contrast CNR on the X-ray energy at three different magnifications in the setup corresponding to Tables 1 and 2 can be found in Fig. 7. In that case, the X-ray energy $E \cong 27$ keV was found to be the optimal one, with the corresponding bulk transmission around 6.5%. The theoretical optimal transmission of 2% is achieved in this case at $E \cong 21$ keV. The discrepancy between the theoretical and numerical results can be attributed to the approximate nature of the assumed dependencies of the linear attenuation coefficients on the X-ray energy (wavelength).

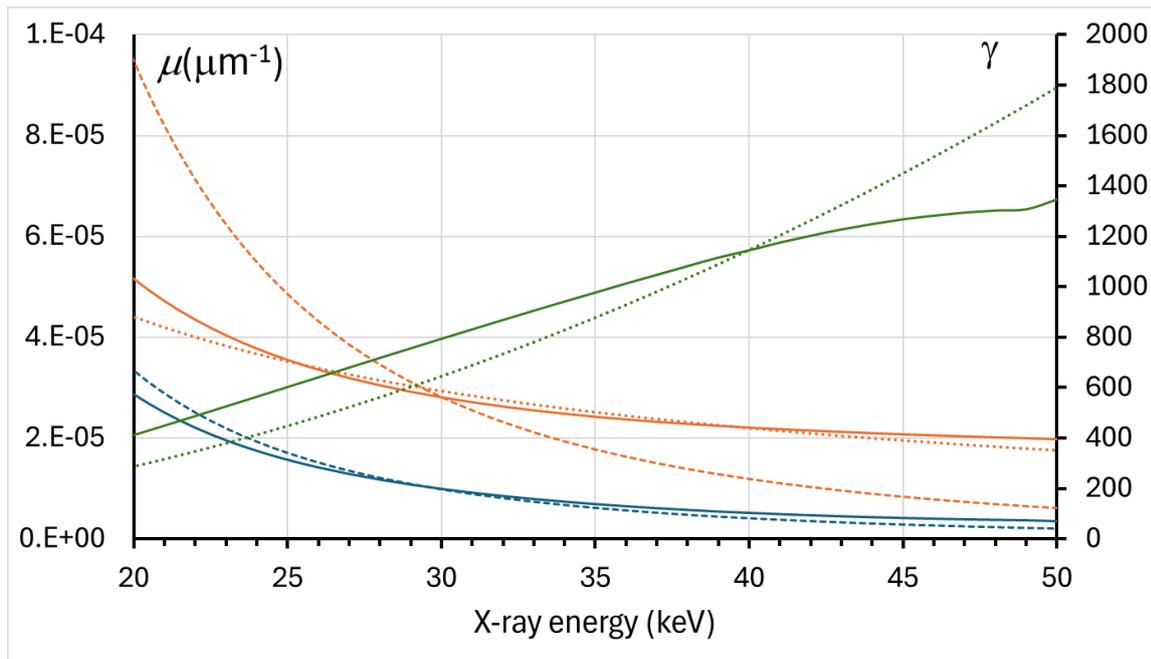

**Figure 6** Linear attenuation coefficient for adipose tissue as a function of X-ray energy ($\mu_0(E)$, solid orange line), with a $E^{-1}$ ($\lambda$-linear) fit (dotted orange line) and a $E^{-3}$ fit (dashed orange line); the difference between linear absorption coefficients for glandular and adipose tissue as a function of X-ray energy ($\mu(E)$, solid blue line), with a $E^{-3}$ fit (dashed blue line); coefficient $\gamma(E)=[\delta_{\text{gland}}(E)-\delta_{\text{adipose}}(E)]/[\beta_{\text{gland}}(E)-\beta_{\text{adipose}}(E)]$ (solid green line), with a $E^2$ fit (dotted green line).

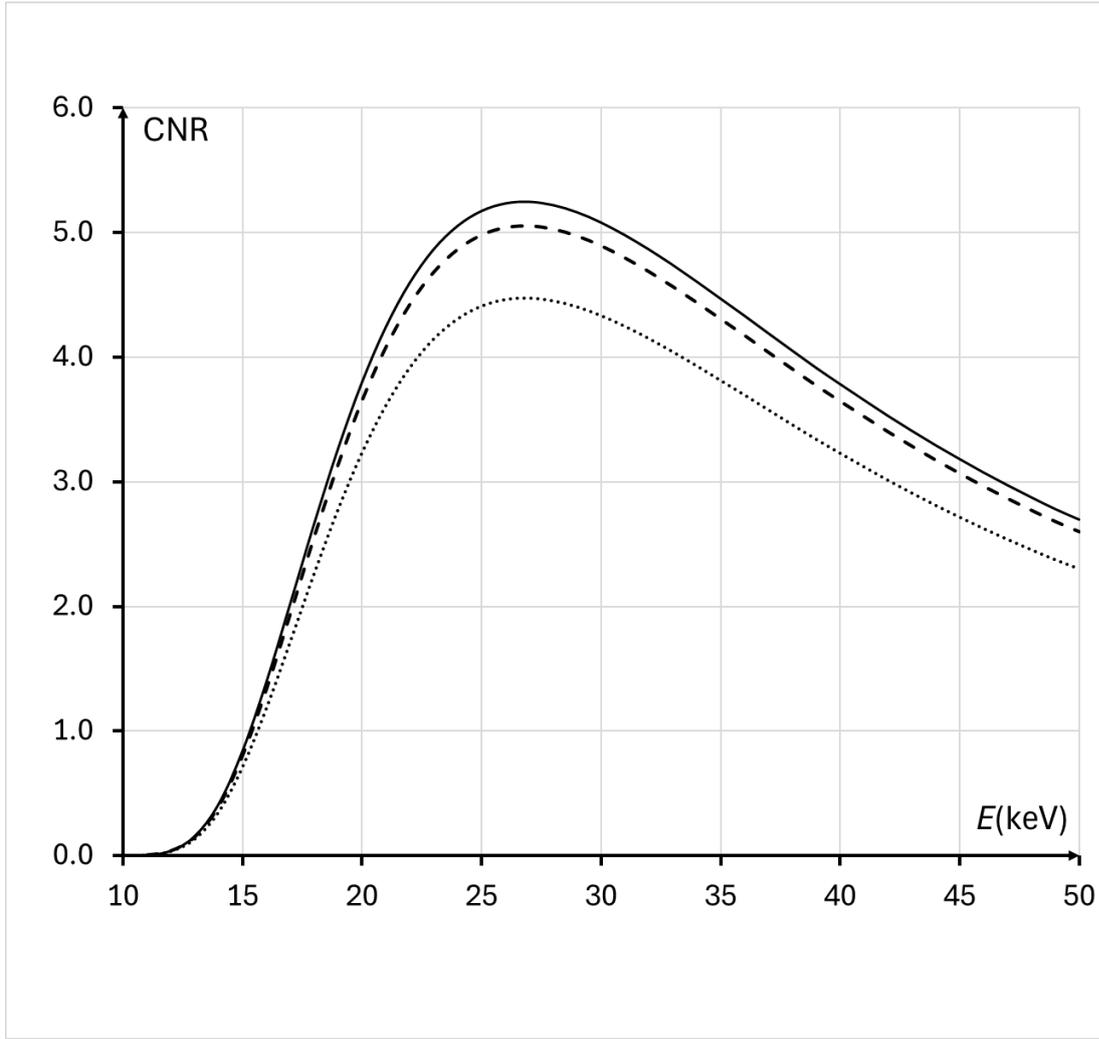

**Figure 7** CNR, as expressed by eq.(13), in the cases corresponding to imaging configurations with parameters from Tables 1 and 2, 10 keV $\leq E \leq$ 50 keV, $I_{in}(\lambda) = 1$ µm$^{-2}$, and different magnifications: $M = 1.15$ ($R_2 \cong 18.3$ m, dotted line), $M = 1.087$ ($R_2 \cong 11.2$ m, solid line), $M = 1.066$ ($R_2 \cong 8.67$ m, dashed line). The optimal energy in this case was $E \cong 27$ keV, at which the average transmission through the sample was ~6.5 %.

## 4. Optimization of biomedical X-ray imaging quality in 2D and 3D

While CNR includes image noise in addition to contrast, it is often essential, especially in biomedical imaging applications, to also take into account the effects of the spatial resolution and the radiation dose delivered to the sample when evaluating X-ray imaging quality. For example, it is possible to argue that the apparent decrease of SNR with magnification in eq.(13) is "superficial", because it corresponds to the reduction of the effective size of the detector pixel as a function of magnification. One could easily apply a low-pass filter that would bring the effective pixel size back to the level corresponding to $M = 1$, which would increase the SNR in proportion to the increased effective pixel size and thus remove the effect of magnification on SNR in this particular respect. For the purpose of properly balancing the essential image quality characteristics, including the CNR, the spatial resolution and the dose, we recently introduced a new metric, termed the "biomedical X-ray imaging quality characteristic" (Gureyev *et al.*, 2025). In the case of 2D imaging, the biomedical X-ray imaging quality characteristic $Q_\mathrm{C}$ was defined as follows (Gureyev *et al.*, 2025):

$$Q_{\mathrm{C,2D}}(M,\lambda) = \frac{\mathrm{CNR}(M,\lambda)\, R_{\mathrm{ab,air}}^{1/2}(\lambda_0)}{\Delta_{\mathrm{sys}} D_{\mathrm{ab}}^{1/2}(\lambda)}, \qquad (14)$$

where $\Delta_{\mathrm{sys}} = 2\sqrt{\pi}\,\sigma_{\mathrm{sys}}(M)$ is the spatial resolution of the imaging system, $D_{\mathrm{ab}}(\lambda) = R_{\mathrm{ab,material}}(\lambda) I_{\mathrm{in}}(\lambda)$ is the absorbed dose (Bezak *et al.*, 2021), $R_{\mathrm{ab,material}}(\lambda) = (\mu_{\mathrm{en}}/\rho)_{\mathrm{material}}(\lambda) E_{\mathrm{ph}}(\lambda)$, $(\mu_{\mathrm{en}}/\rho)_{\mathrm{material}}$ is the mass energy-absorption coefficient of a given material at wavelength $\lambda$, $\lambda_0$ is a fixed wavelength corresponding to a particular X-ray energy at which the normalization coefficient $R_{\mathrm{ab,air}}^{1/2}(\lambda_0)$ is evaluated, and $E_{\mathrm{ph}}(\lambda) = hc/\lambda$ is the energy of a single photon, $h$ is the Planck constant and $c$ is the speed of light (Hubbell & Seltzer, 1996). We fixed $\lambda_0$ in the numerator of eq.(14), thus slightly modifying the definition of the biomedical X-ray imaging quality introduced in Part 2 of this paper, in order to properly optimize the imaging conditions with respect to the absorbed dose. Indeed, the optimization should be performed with the goal of minimising the "absolute" dose absorbed by the sample, rather than the sample dose relative to dose to air at the same X-ray energy. The fact that air may be absorbing a smaller or a larger dose at different X-ray energies is irrelevant to the task of minimization of the dose delivered to the sample.

Another subtle difference between eq.(14) and the biomedical X-ray imaging quality introduced in (Gureyev *et al.*, 2025) is in the definition of the contrast. Equation (14) includes the propagation contrast defined in eq.(9) above for our simple model of an embedded edge. On the other hand, the more general formulation of $Q_{C,2D}(M,\lambda)$ uses the contrast defined as a ratio of the difference between the average values of image intensity in two adjacent regions, divided by the maximum of the two average values (Gureyev *et al.*, 2025). In this context, choosing a suitable definition of contrast depends on the selected optimization task. In the case of PBI of an edge feature in a near-Fresnel region, CNR is described by eq.(13). As we are mostly interested in phase contrast produced by weakly absorbing samples, we shall neglect the (typically, small) term corresponding to absorption contrast in eq.(13) and consider only the phase-contrast term in the case of a sharp weakly-absorbing edge. Substituting the latter term into eq.(14), we obtain:

$$Q_{C,2D}(M,\lambda) \cong \frac{\eta^{1/2}(\mu T)_{max}(\lambda)\exp[-(\mu T)_{max}(\lambda)-\mu_0(\lambda)T_0/2]}{(2\pi e)^{1/2}K^{1/2}(\lambda,\lambda_0)} \frac{\Delta_{det}\gamma}{M\Delta_{sys}N_F}, \qquad (15)$$

where $K(\lambda,\lambda_0) = R_{ab,material}(\lambda)/R_{ab,air}(\lambda_0) = (\lambda_0/\lambda)(\mu_{en}/\rho)_{material}(\lambda)/(\mu_{en}/\rho)_{air}(\lambda_0)$. The "dose conversion coefficient" $K(\lambda,\lambda_0)$ reflects the behaviour of the mean X-ray dose absorbed by the feature relative to the entrance air kerma (Bezak *et al.*, 2021) at a particular X-ray energy $E_0 = hc/\lambda_0$. The choice of this energy is unimportant, since the factor $R_{ab,air}(\lambda_0)$ is included in the expression for $Q_{C,2D}(M,\lambda)$ only for the purpose of normalization and making the quantity dimensionless (Gureyev *et al.*, 2025). The asymmetry in the roles of $\Delta_{src}$ and $\Delta_{det}$ in eq.(15) reflects the fact that the source size and the detector resolution affect the imaging quality in different ways: the source size contributes to the spatial resolution similarly to the detector resolution, but, unlike the detector resolution, does not contribute to the SNR. When $M = 1$ and hence the source size does not affect the image, we have $\Delta_{det}/(M\Delta_{sys}) = 1$.

As in the case of contrast above, we shall consider the problems of optimization (maximization) of the biomedical X-ray imaging quality as a function of magnification and the X-ray wavelength. As in the case of PBI contrast, apart from the constant factor $\eta^{1/2}/(2\pi e)^{1/2}$, the biomedical image quality factorizes into a product of two distinct terms, $R'\Delta_{det}/(M\Delta_{sys}^3)$ and $\gamma(\lambda)\lambda(\mu T)_{max}(\lambda)\exp[-(\mu T)_{max}(\lambda)-\mu_0(\lambda)T_0/2]K^{-1/2}(\lambda,\lambda_0)$, the first one being a function of the geometrical parameters of the imaging setup and the second one depending on the X-ray wavelength.

As above, we consider the case of a fixed total source-to-detector distance $R = R_1 + R_2$, where the expression $R'\Delta_{det}/(M\Delta_{sys}^3)$ needs to be maximized as a function of magnification. Expressing $R'\Delta_{det}/(M\Delta_{sys}^3) = (4\pi)^{-1}R(M-1)\sigma_{det}/[(M-1)^2\sigma_{src}^2+\sigma_{det}^2]^{3/2}$, it is easy to check that the equation $d[R'\Delta_{det}/(M\Delta_{sys}^3)]/d(M-1) = 0$ has the solution $M_{Q2} = 1+\sigma_{det}/(\sqrt{2}\sigma_{src})$. At this optimal magnification we have $[R'\Delta_{det}/(M\Delta_{sys}^3)](M_{Q2}) = 2R/(\sqrt{27}\Delta_{det}\Delta_{src})$. Therefore, the biomedical X-ray imaging quality in PBI of a sharp monomorphous edge, corresponding to the magnification $M_{Q2}$, is equal to

$$Q_{C,2D}(M_{Q2},\lambda) = \frac{\eta^{1/2}(\mu T)_{max}(\lambda)\exp[-(\mu T)_{max}(\lambda)-\mu_0(\lambda)T_0/2]}{[(27/2)\pi e K(\lambda)]^{1/2}}\frac{\gamma R\lambda}{\Delta_{src}\Delta_{det}}, \quad (16)$$
$$M_{Q2} = 1+\sigma_{det}/(\sqrt{2}\sigma_{src}).$$

The biomedical X-ray imaging quality in eq.(16) is linearly proportional to the total source-to-detector distance and is inversely proportional to both the source size and the detector resolution. When $\sigma_{det} = \sigma_{src}$, we obtain $M_{Q2} \cong 1.707$, while in the case corresponding to the IMBL imaging setup parameters in Tables 1 and 2, $M_{Q2} \cong 1.066$ ($R_2 \cong 8.7$ m). The latter result agrees with direct numerical evaluation of eq.(15) presented in Fig. 8 for an imaging setup corresponding to Tables 1 and 2.

Regarding the optimization of $Q_{C,2D}(M,\lambda)$ with respect to $\lambda$, we first note that at hard X-ray energies, $20\,\text{keV} \le E \le 50\,\text{keV}$, the mass energy-absorption coefficient of soft biological tissues is

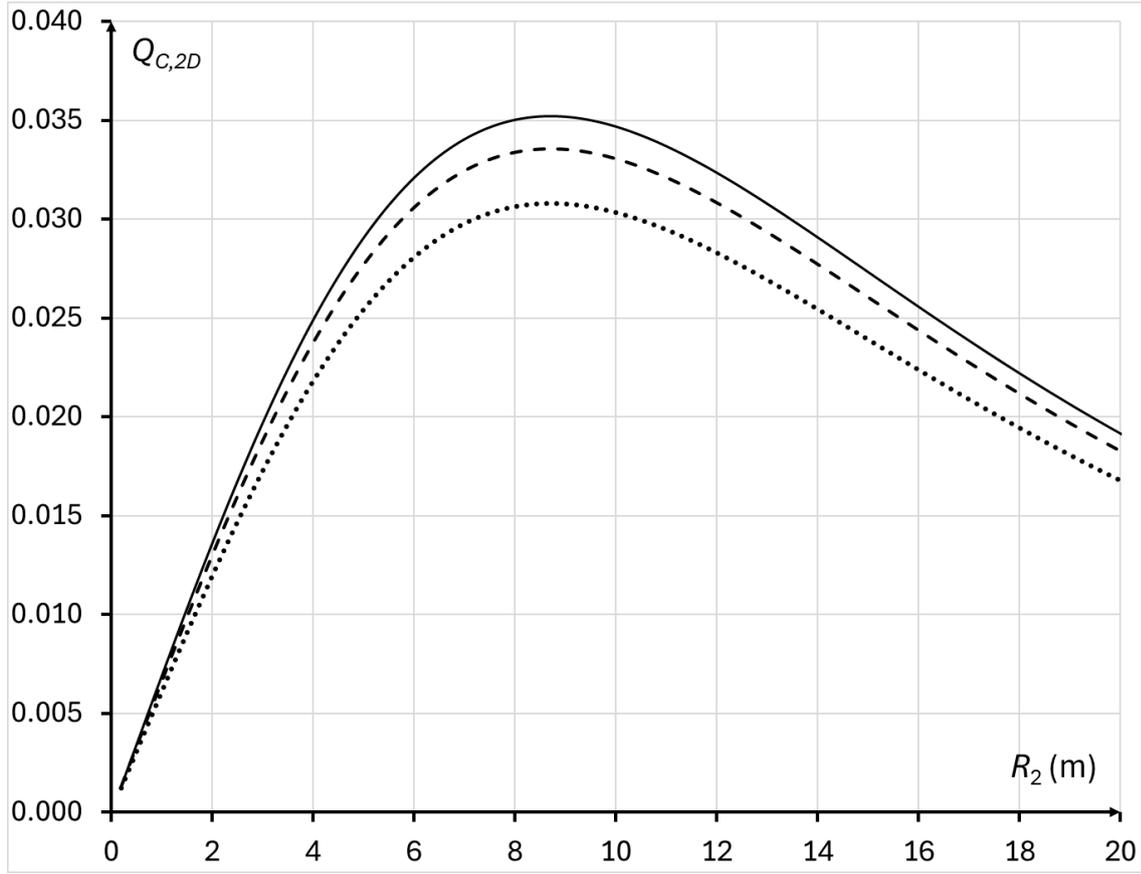

**Figure 8** Biomedical X-ray imaging quality $Q_{C,2D}(M_{Q2}, \lambda)$, as expressed by eq.(15), in the cases corresponding to some proposed configurations for imaging breast tissue specimens at the IMBL beamline of the Australian Synchrotron (see Tables 1 and 2), with 0.1 m ≤ $R_2$ ≤ 20 m, $I_{in}(\lambda) = 1$ μm$^{-2}$, and different X-ray wavelengths (energies): $\lambda$ = 0.4769 Å ($E$ = 26 keV, dotted line), $\lambda$ = 0.3875 Å ($E$ = 32 keV, solid line), $\lambda$ = 0.2952 Å ($E$ = 42 keV, dashed line). The optimal magnification in this case is equal to $M_Q \cong 1.066$ ($R_2 \cong 8.7$ m).

expected to be approximately proportional to the third power of the wavelength, similarly to the linear attenuation coefficient (Chantler *et al.*, 1997; NIST, 2025). As a consequence, the coefficient $K(\lambda, \lambda_0) = const(\lambda_0)\lambda^{-1}(\mu_{en}/\rho)_{material}(\lambda)$ is approximately proportional to $\lambda^2$, i.e. $K(\lambda, \lambda_0) \cong K(\lambda_0, \lambda_0)(\lambda/\lambda_0)^2$. Using the wavelength dependencies already considered above for the other quantities in eq.(16), we can again introduce a temporary notation here:

$\gamma(\lambda)\lambda K^{-1/2}(\lambda, \lambda_0) \cong \gamma(\lambda_0)\lambda_0 K^{-1/2}(\lambda_0, \lambda_0)(\lambda_0/\lambda)^{-2} = a\lambda^{-2}$,

$(\mu T)_{max}(\lambda) = (\mu T)_{max}(\lambda_0)(\lambda/\lambda_0)^3 = b\lambda^3$ and

$(\mu T)_{max}(\lambda) + \mu_0(\lambda)T_0/2 \cong \mu_0(\lambda_0)(T_0/2)(\lambda/\lambda_0) = c\lambda$. Then we need to find a maximum of the function $g(\lambda) = \gamma(\lambda)\lambda K^{-1/2}(\lambda, \lambda_0)(\mu T)_{max}(\lambda)\exp[-\mu_0(\lambda)T_0/2] = ab\lambda\exp(-c\lambda)$. The equation $dg(\lambda)/d\lambda = ab(1-c\lambda)\exp(-c\lambda) = 0$ has a root $\lambda_{Q2} = 1/c$, which corresponds to the maximum $g(\lambda_{Q2}) = e^{-1}(\mu T)_{\max}(\lambda_{Q2})\gamma(\lambda_{Q2})\lambda_{Q2}K^{-1/2}(\lambda_{Q2},\lambda_0)$. In practice, when the edge feature is small compared to the bulk object, one has $(\mu T)_{\max}(\lambda) \ll \mu_0(\lambda)T_0/2$, and the optimal wavelength $\lambda_{Q2}$ can be found experimentally from the condition $\exp[-\mu_0(\lambda_{Q2})T_0] = e^{-2} \cong 0.14$, corresponding to the requirement that the mean X-ray transmission through the sample should be around 14%.

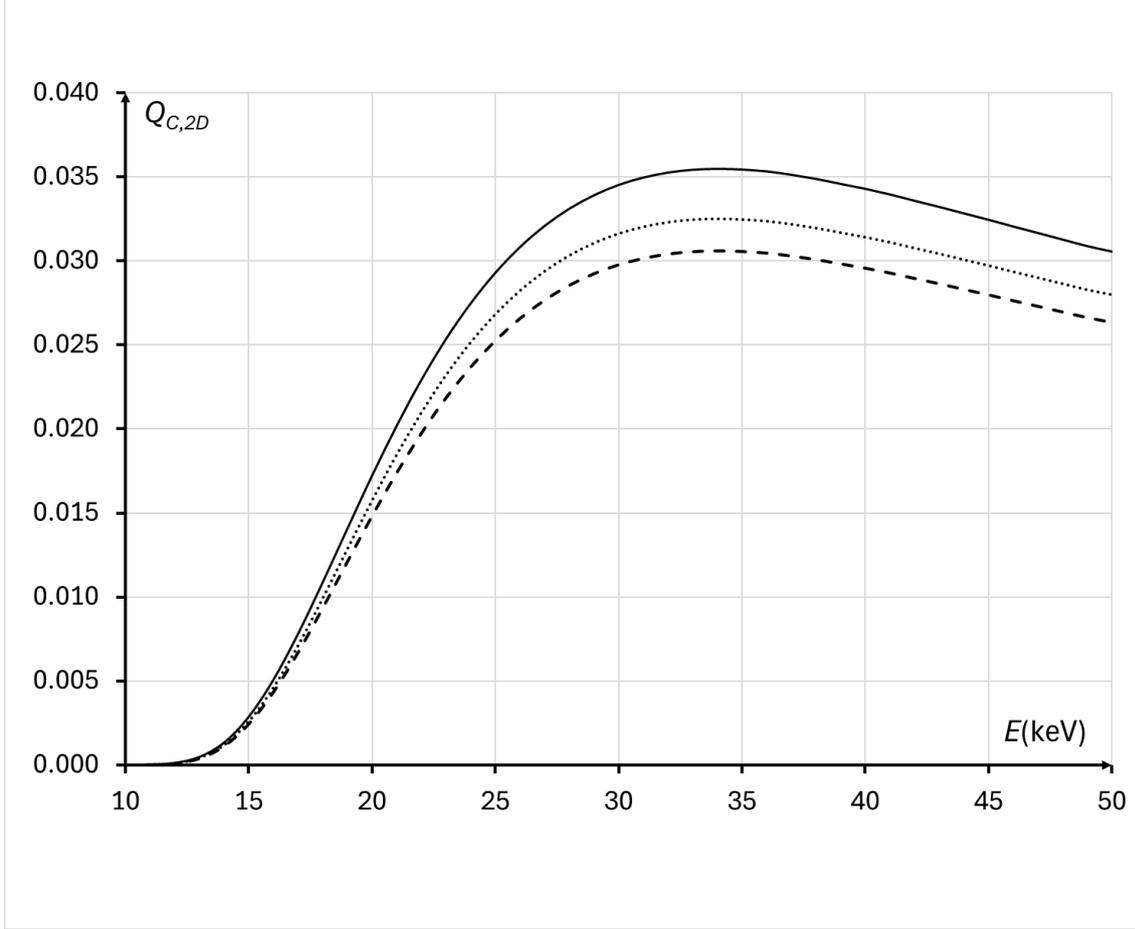

**Figure 9** Biomedical X-ray imaging quality $Q_{C,2D}(M,\lambda)$, as expressed by eq.(15), in the cases corresponding to imaging configurations with parameters from Tables 1 and 2, 10 keV $\leq E \leq$ 50 keV, $I_{in}(\lambda) = 1$ μm$^{-2}$, and different magnifications: $M = 1.094$ ($R_2 \cong 12.03$ m, dotted line), $M = 1.066$ ($R_2 \cong 8.67$ m, solid line), $M = 1.04$ ($R_2 \cong 5.38$ m, dashed line). The optimal energy in this case was $E \cong 34$ keV.

We have also performed direct numerical evaluations of eq.(15), using the imaging setup parameters from Tables 1 and 2, within the range of X-ray energies $10\,\text{keV} \leq E \leq 50\,\text{keV}$ and at three different magnifications: $M = 1.066$, $M = 1.094$, $M = 1.04$ (Fig. 9). These calculations confirmed that, at all considered X-ray energies, the best magnification was $M = 1.066$, in agreement with the theoretical optimization result presented above. The same calculations also showed that the optimal energy maximizing $Q_{C,2D}(M,\lambda)$ was $E \cong 34$ keV ($\lambda \cong 0.3263$ Å). The average transmission through the bulk of the sample at that energy was approximately 12.0 %, which was slightly lower than the predicted optimal transmission of 13.5 %, corresponding to the energy of 37 keV. Note however that the difference between the values of $Q_{C,2D}(M,\lambda)$ at 34 keV and 37 keV was less than 1%.

Finally, we considered the problem of optimization of the 3D biomedical X-ray imaging quality of PB-CT. The following expression can be easily derived from the corresponding result in the parallel-beam case found in (Gureyev *et al.*, 2025):

$$Q_{C,3D}(M,\lambda) \cong \frac{\sqrt{12}\,\eta^{1/2}(\mu L)\exp(-\mu_0 L/2)}{\pi K^{1/2}(\lambda,\lambda_0)} \frac{\Delta_{det}^2 \gamma}{M\,L^{1/2}\Delta_{sys}^{3/2} N_F} f(M,\lambda), \tag{17}$$

where $\mu = (\mu_1 - \mu_0) > 0$, $L = \pi R_{CT}/2$, $R_{CT}$ is the radius of the cylindrical volume of the CT reconstruction and $f(M,\lambda) = (\pi/6)^{1/2}[\ln(\gamma/N_F)-1]^{-1/2}$. We will consider the case of relatively large Fresnel numbers, where $\ln(\gamma/N_F) \gg 1$. In such cases, the term $f(M,\lambda)$ is slowly varying and can be neglected in an analytical optimization. However, we will still include the factor $f(M,\lambda)$ in the direct numerical evaluation of eq.(17) used for comparison with the analytical results below. Note that in the context of eq.(17), the feature of interest is no longer limited to the blurred monomorphous edge model used above. However, both the imaged sample and the feature of interest are still assumed to be approximately monomorphous (Gureyev *et al.*, 2025). Another difference with the 2D imaging case considered above is in the fact that eq.(17) utilises the image contrast $C_m = (\mu_1 - \mu_0)/\mu_1$ defined as a ratio of the difference between the average values of X-ray attenuation in the reconstructed feature of interest and its surroundings (background), divided by the attenuation in the feature of interest (Gureyev *et al.*, 2025).

Equation (17) uses one variant of the 3D "gain coefficient" obtained in (Nesterets & Gureyev, 2014). We also performed the optimizations with a different variant of eq.(17) containing an alternative expression for the 3D gain coefficient (Gureyev *et al.*, 2025), which led to very similar results for $Q_{C,3D}(M,\lambda)$, with a difference of about 10% that was nearly uniform across the tested range of propagation distances and energies.

As above, we consider first the case of a fixed total source-to-detector distance $R = R_1 + R_2$, where the expression $F(M) = R'/(M\Delta_{sys}^{7/2}) = (2\sqrt{\pi})^{-7/2} RM^{1/2}(M-1)/[(M-1)^2\sigma_{src}^2 + \sigma_{det}^2]^{7/4}$ needs to be maximized as a function of $M$. The equation $dF/d(M-1) = 0$ can be reduced to the cubic equation $4(M_{Q3}-1)^3 + 5(M_{Q3}-1)^2 - 3(\sigma_{det}^2/\sigma_{src}^2)(M_{Q3}-1) - 2(\sigma_{det}^2/\sigma_{src}^2) = 0$ with respect to the unknown value of the optimal magnification $M_{Q3}$. Roots of this equation can be expressed analytically in terms of the quantity $\sigma_{det}^2/\sigma_{src}^2$ using Cardano's formula, but the corresponding expressions are not very useful. It is also possible to use Wolfram Mathematica (Wolfram Research Inc., 2025) or similar tools for this purpose. Finally, rewriting the cubic equation in the form $\sigma_{det}/\sigma_{src} = (M_{Q3}-1)[(4M_{Q3}+1)/(3M_{Q3}-1)]^{1/2}$ provides a one-to-one correspondence between the optimal magnification values and the corresponding ratios of the detector resolution to the source size that can be used as a look-up table (see the dashed line in Fig. 5). It can be seen from Fig. 5 that $M_{Q3} < M_C = 1 + \Delta_{det}/\Delta_{src}$ for all values of $\Delta_{det}/\Delta_{src}$, and $M_{Q3} \cong M_{Q2} = 1 + \Delta_{det}/(\sqrt{2}\Delta_{src})$. In the case $\sigma_{src} = \sigma_{det}$, we obtain $M_{Q3} \cong 1.727$ ($M_{Q2} \cong 1.707$ in this case). In the case corresponding to imaging setup parameters in Tables 1 and 2, the positive root of the cubic equation is $M_{Q3} \cong 1.060$ ($R_2 \cong 7.9$ m). This value is rather close to $M_{Q2} \cong 1.066$ ($R_2 \cong 8.7$ m) in the same case. A direct numerical evaluation of eq.(17) with the same parameters gives the optimum magnification value $M_{Q3} \cong 1.053$ ($R_2 \cong 7.0$ m) (Fig. 10). The difference between the analytical and numerical results here is likely due to the fact (mentioned above) that the analytical optimization did not take into account the slowly varying factor $f(M,\lambda)$ in eq.(17). When we used the alternative variant of eq.(17) containing the 3D gain coefficient from (Nesterets & Gureyev, 2014) for direct numerical evaluation of the biomedical X-ray imaging quality, the optimum magnification became $M_{Q3} \cong 1.055$ ($R_2 \cong 7.3$ m). The differences between the values of the biomedical X-ray imaging quality in the configurations with $R_2 \cong 7.3$ m, 7.0 m and 7.9 m were very small, because $Q_{C,3D}$ changes slowly near the point of maximum (see e.g. Fig. 10). Note that the optimum magnification $M_{Q3}$ is also independent of the wavelength $\lambda$, i.e. it is the same for any X-ray energy.

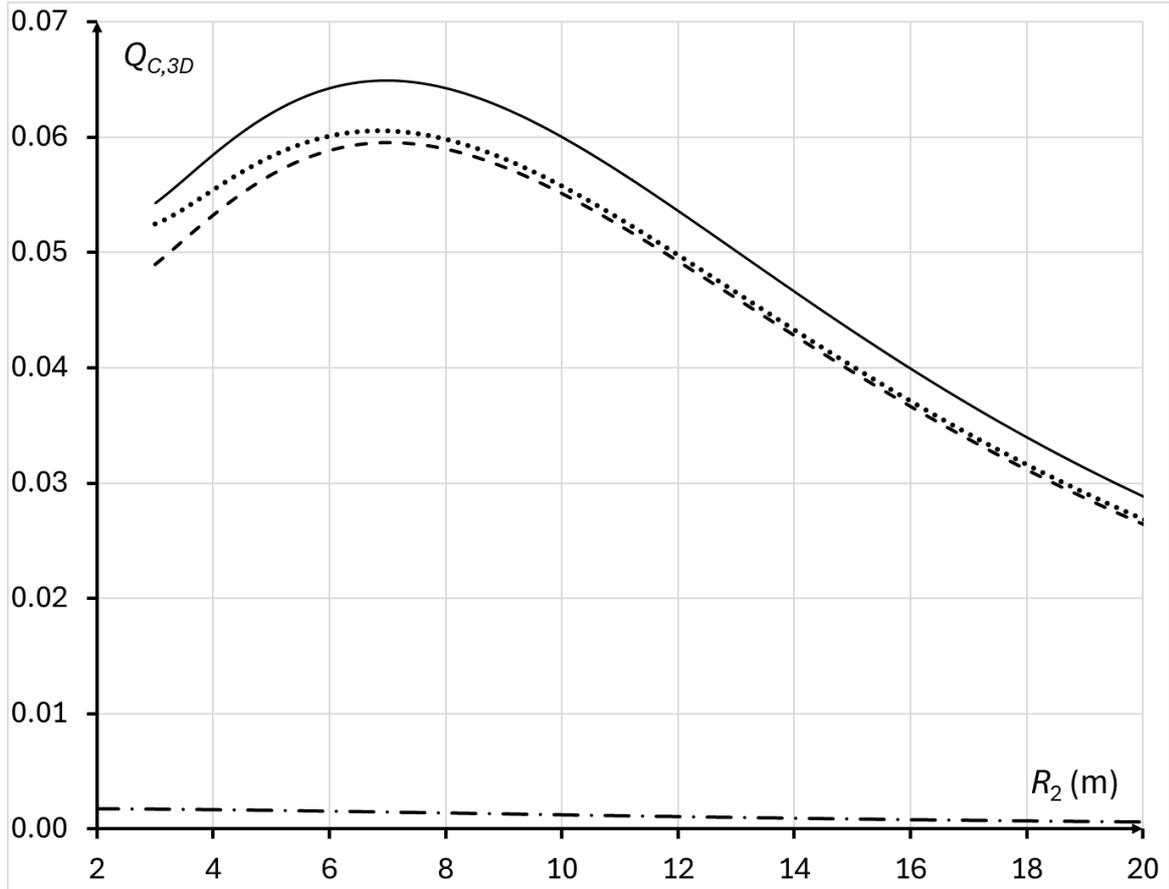

**Figure 10** Biomedical X-ray imaging quality $Q_{C,3D}(M,\lambda)$, as expressed by eq.(17), in the cases corresponding to imaging configurations with parameters from Tables 1 and 2, 2 m ≤ $R_2$ ≤ 20 m, $I_{in}(\lambda)$ = 1 µm$^{-2}$, and different X-ray wavelengths (energies): $\lambda$ = 0.4769 Å ($E$ = 26 keV, dotted line), $\lambda$ = 0.3875 Å ($E$ = 32 keV, solid line), $\lambda$ = 0.2952 Å ($E$ = 42 keV, dashed line). The dot-dash line shows the biomedical X-ray imaging quality for the pure absorption case at $E$ = 32 keV. The optimal magnification in this case was equal to $M_{Q3} \cong 1.053$ ($R_2 \cong 7.0$ m).

Regarding the optimization of $Q_{C,3D}(M,\lambda)$ with respect to $\lambda$, we follow the same approach as used above for $Q_{C,2D}(M,\lambda)$ and $C(M,\lambda)$. We previously established that

$K(\lambda,\lambda_0) \cong K(\lambda_0,\lambda_0)(\lambda/\lambda_0)^2$, $\mu(\lambda) \cong \mu(\lambda_0)(\lambda/\lambda_0)^3$, $\mu_0(\lambda) \cong \mu_0(\lambda_0)(\lambda/\lambda_0)$, and $\gamma(\lambda)\lambda \cong \gamma(\lambda_0)\lambda_0(\lambda/\lambda_0)^{-1}$. In the case of eq.(17) we need to find a maximum of the function $h(\lambda) = \gamma(\lambda)\lambda K^{-1/2}(\lambda)(\mu L)(\lambda)\exp[-(\mu_0 L)(\lambda)/2] = a\lambda\exp(-b\lambda)$. The equation $dh(\lambda)/d\lambda = a(1-b\lambda)\exp(-b\lambda) = 0$ has a root $\lambda_{Q3} = 1/b$, which corresponds to a maximum, $h(\lambda_{Q3}) = e^{-1}\gamma(\lambda_{Q3})\lambda_{Q3}K^{-1/2}(\lambda_{Q3})(\mu L)(\lambda_{Q3})$. In practice, the optimal wavelength $\lambda_{Q3}$ can be found experimentally from the condition $\exp[-\mu(\lambda_{Q3})L] = e^{-2} \cong 0.135$, corresponding to the requirement that the mean X-ray transmission through the sample should be around 13.5 %.

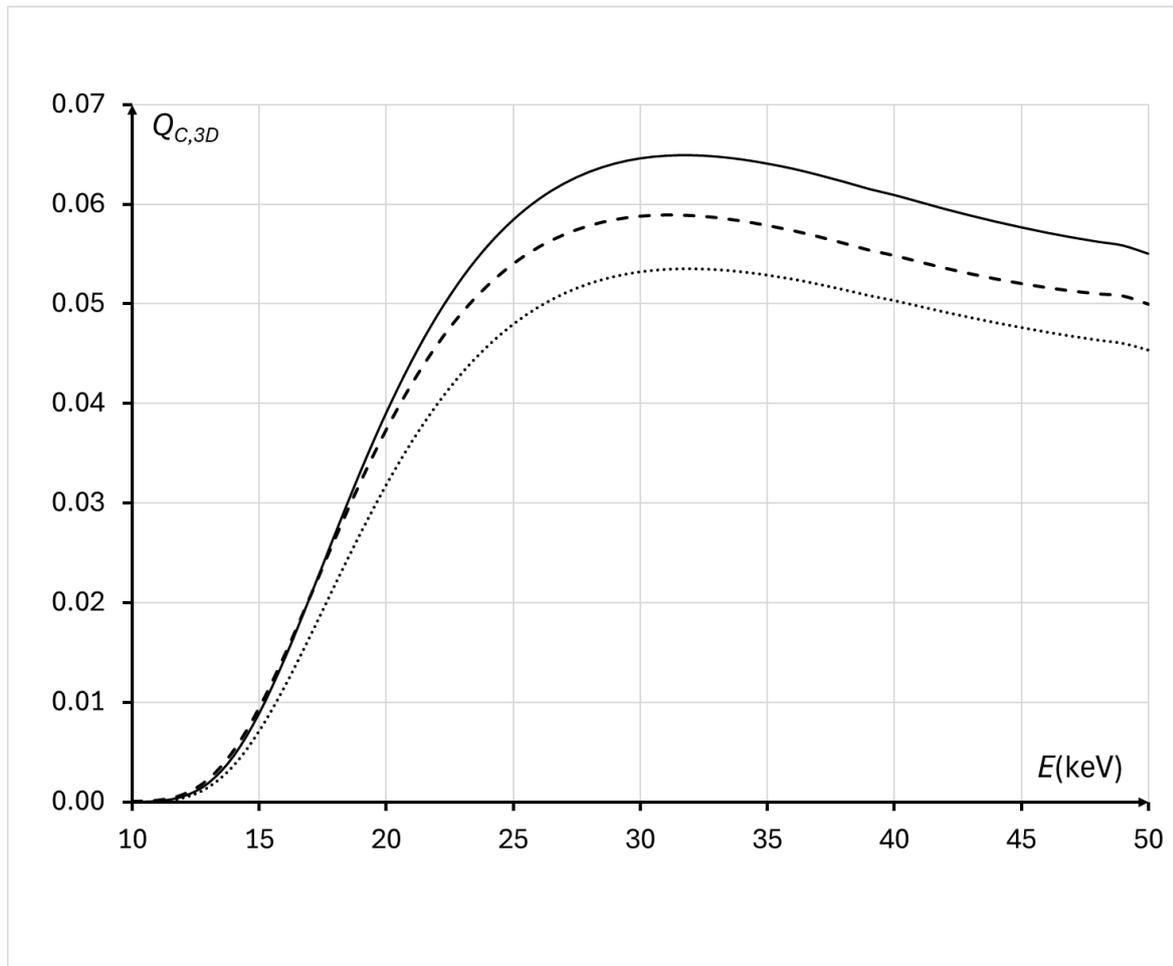

**Figure 11**   Biomedical X-ray imaging quality $Q_{C,3D}(M,\lambda)$, as expressed by eq.(17), in the cases corresponding to imaging configurations with parameters from Tables 1 and 2, $10\,\text{keV} \leq E \leq 50\,\text{keV}$, $I_{\text{in}}(\lambda) = 1\,\mu\text{m}^{-2}$, and different magnifications: $M = 1.094$ ($R_2 \cong 12.03$ m, dotted line), $M = 1.053$ ($R_2 \cong 7.0$ m, solid line), $M = 1.03$ ($R_2 \cong 4.08$ m, dashed line). The optimal energy in this case was $E \cong 32$ keV.

We have also performed direct numerical evaluation of eq.(17), using the imaging setup parameters from Tables 1 and 2, within the range of X-ray energies $10\,\text{keV} \leq E \leq 50\,\text{keV}$ and at three different magnifications: $M = 1.053$, $M = 1.094$, $M = 1.03$ (Fig. 11). These calculations confirmed that, for all X-ray energies, the best magnification was $M = 1.053$, in agreement with the theoretical optimization results presented above. The same calculations also showed that the optimal energy maximizing $Q_{C,3D}(M,\lambda)$ was $E \cong 32$ keV ($\lambda \cong 0.3351$ Å). The average transmission through the bulk of the sample at that energy was approximately 10.5 %, which was lower than the predicted optimal transmission of 13.5 %, which corresponded to the energy of 37 keV. Note however that the difference between the values of $Q_{C,3D}(M,\lambda)$ at 32 keV and 37 keV was only 3.0 %.

## 5. Conclusions

We have derived simple analytical expressions for the contrast and spatial resolution in propagation-based phase-contrast images of a model corresponding to a homogeneous edge feature inside a uniform sample. These expressions explicitly show the dependence of the image characteristics on the geometrical parameters of the imaging setup (the source size, the detector resolution, the source-to-sample and the sample-to-detector distances) and on the X-ray wavelength. These explicit dependencies made it possible to perform analytical optimization of the spatial resolution, the contrast and the biomedical X-ray imaging quality characteristics $Q_{C,2D}$ and $Q_{C,3D}$ with respect to the geometric parameters of the setup and the X-ray wavelength. The results of this optimization using eqs.(7)-(17) demonstrate some intuitively expected and physically meaningful features. In the case of CNR and biomedical X-ray imaging quality characteristics, the optimal X-ray wavelength corresponded to transmission of the order of 10% through the bulk of the sample. This reflects a balance between maximization of the image contrast through stronger absorption and phase shifts in the feature of interest, and the need to still obtain a sufficiently strong SNR at the detector plane, which gets weaker when more photons are absorbed in the bulk of the sample.

The contrast and CNR in PBI increase linearly with the source-to-detector distance within the near-Fresnel region. For a fixed total source-to-detector distance, the behaviour of these characteristics is less straightforward with respect to the geometric magnification, i.e. as a function of the ratio of the source-to-detector and source-to-sample distances. The optimal magnification is determined by the ratio of the X-ray source size and the detector resolution. In the case of quantities that do not depend on the image noise and the radiation dose, such as spatial resolution and contrast, the optimal configurations are symmetric with respect to detector resolution and source size. At the optimal magnification, these image quality characteristics are inversely proportional to the product of the source size and the detector resolution. In other words, at the optimal magnification, the blurring due to the source size and the detector PSF contribute equally to the image. On the other hand, quantities such as CNR, $Q_{C,2D}$ and $Q_{C,3D}$ − which take into account the photon shot noise and the radiation dose, in addition to the contrast and spatial resolution – no longer exhibit such symmetry. In other words, inclusion of photon noise into the quality metrics breaks the symmetry between the contributions of the source size and the detector resolution. This happens because, while the increased blurring due to broader PSF of the detector proportionally increases the SNR (in accordance with the noise-resolution duality (Gureyev *et al*., 2014, 2016)), the increase of the penumbral blurring due to the X-ray source size does not lead to an increase of the SNR. The latter fact is a consequence of the nature of typical X-ray sources, including fixed-anode microfocus sources and synchrotron sources based on present-day insertion devices such as wigglers and undulators. Such sources can be modelled as a collection of independent point-like radiators, as in the case of classical thermal sources (Pelliccia & Paganin, 2025). As a result, the photons reaching the detector from different parts of the source are statistically independent. This lack of spatial photon correlation, and the consequential absence of any increase in the SNR related to the source size (provided that the photon fluence remains constant), is in contrast with the correlations induced by convolution with the detector PSF (Goodman, 2000). The asymmetry in the effects of the source size and the detector resolution on the image noise reduces the optimal magnification values, suppressing the source size more than the detector resolution at the optimal magnification.

**Table 3** Summary of optimal magnifications and energies that maximize various image quality metrics in PBI.

|  | $M_{opt}$ | $E_{opt}$ |
|---|---|---|
| **Resolution** | $1+(\sigma_{det}^2/\sigma_{src}^2)$ | N/A |
| **Contrast** | $1+(\sigma_{det}/\sigma_{src})$ | $\exp[-(\mu T)_{max}(E_{opt})] \cong 0.51$ |
| **CNR** | $\sigma_{det}/\sigma_{src} = (M_{opt}-1)(2M_{opt}-1)^{1/2}$ | $\exp[-\mu_0(E_{opt})T_0] \cong 0.02$ |
| **Q$_{C,2D}$** | $1+[\sigma_{det}/(\sqrt{2}\sigma_{src})]$ | $\exp[-\mu_0(E_{opt})T_0] \cong 0.14$ |
| **Q$_{C,3D}$** | $\sigma_{det}/\sigma_{src} = (M_{opt}-1)[(4M_{opt}+1)/(3M_{opt}-1)]^{1/2}$ | $\exp[-\mu_0(E_{opt})T_0] \cong 0.14$ |

The results of PBI optimization with respect to the geometrical magnification (sample-to-detector distance $R_2$) presented in this paper are rather straightforward and accurate, as they are based on precise mathematical dependencies on the relevant geometric parameters. In contrast, our optimisation with respect to the X-ray wavelength (energy) involved relatively crude approximations for the functional dependencies of factors like the complex refractive index of materials on the X-ray energy. Therefore, the latter results are likely to be less broadly applicable in their current form. In practice, it may be preferable to carry out optimizations with respect to the X-ray energy for a given imaging setup by using the analytical expressions derived in the present paper in combination with tabulated values of the complex refractive index and the mass energy-absorption coefficient as functions of the X-ray energy (Hubbell & Seltzer, 1996; Chantler *et al.*, 1997).

We have performed direct numerical evaluation of the obtained analytical expressions for the contrast, CNR, $Q_{C,2D}$ and $Q_{C,3D}$, for a set of parameters that approximately correspond to current and prospective setups for imaging breast tissue specimens at IMBL (Gureyev *et al.*, 2019) (Figs. 2-4, 7-11). These simulations not only allowed us to verify the relevant analytical results obtained for the optimum imaging conditions, but also provided examples of procedures that can be used for numerical optimization of geometric parameters and X-ray energy under specified experimental conditions. Remarkably, the optimal magnification and the X-ray energy obtained in the calculations for the 3D biomedical X-ray imaging quality characteristic, $Q_{C,3D}$, i.e. $M = 1.032$ ($R_2 = 7$ m) and $E = 32$ keV, agreed quite well with the previously reported optimal values obtained in connection with breast cancer PB-CT imaging work at synchrotron beamlines (Baran et al., 2017; Brombal et al., 2018; Brombal, 2020; Taba et al., 2019; Gureyev *et al*., 2019). Although the optimizations were performed in the present work only for monochromatic X-rays, the obtained results show a clear path towards optimization for polychromatic spectra. Firstly, we have shown that the optimization with respect to magnification and the energy can be performed independently of each other, and, in particular, the optimal magnification remains the same for all X-ray energies within the validity range of the used approximations. Secondly, the simulation results for the energy dependence of $Q_{C,2D}$ and $Q_{C,3D}$ presented in Figs. 9 and 11, indicate that these characteristics change very slowly after the X-ray energy is increased beyond a certain "lower threshold" (approximately 25 keV in the case of breast PBI). Such a conclusion is in line with the general understanding that low-energy X-rays are detrimental to biomedical image quality, because they significantly contribute to the radiation dose, but not to the SNR, as most low-energy photons are absorbed in the sample and do not reach the detector. Once the lower X-ray energies in the spectrum are filtered out, the details of the remaining high-energy spectrum are not going to significantly affect the image quality.

**Table 4**  Summary of optimal magnifications and energies that maximize various image quality metrics under the imaging conditions from Tables 1 and 2 which correspond to existing and prospective configurations for PBI and PB-CT at IMBL (Australian Synchrotron).

|  | $M_{opt}$ | $E_{opt}$ |
|---|---|---|
| **Resolution** | 1.009 ($R_2$ = 1.22 m) | N/A |
| **Contrast** | 1.094 ($R_2$ = 12.0 m) | 12.0 keV |
| **CNR** | 1.087 ($R_2$ = 11.2 m) | 27 keV |
| **$Q_{C,2D}$** | 1.066 ($R_2$ = 8.7 m) | 34 keV |
| **$Q_{C,3D}$** | 1.053 ($R_2$ = 7.0 m) | 32 keV |

We have shared our Excel spreadsheets used for numerical calculations in the present study (Gureyev, 2025). These spreadsheets can be used for similar calculations by inserting suitable values for the geometric parameters of the imaging setup of interest, including the source size, the detector resolution, the X-ray wavelength, as well as the complex refractive index of the sample and some other relevant parameters which can be found in online databases (e.g. NIST, 2025; TS-Imaging, 2025). We hope that these simple spreadsheets can be useful for other researchers in their theoretical and experimental studies involving PBI imaging.

**Data availability**     Excel spreadsheets, including the experimental parameters, used for calculation in this paper are publicly available at https://github.com/timg021/PBI-Optimization/tree/main .